\documentclass[
aps,prb,amsmath,amssymb,
a4paper,twocolumn,notitlepage,longbibliography,superscriptaddress
,footinbib
]{revtex4-2}
\usepackage{graphicx}% Include figure files
\usepackage{dsfont}     % needed for \mathds{1}
\usepackage[usenames,dvipsnames]{xcolor}
\usepackage{xspace}  % needed for comment macros \JS{} etc
\usepackage{braket}  % needed 
\usepackage{bm}        % bold math needed for \Sket{ }
\usepackage{mathtools} % drcases environment
\usepackage{cancel} % to strikethrough
\usepackage{hyperref}
\hypersetup{colorlinks=true,linktoc=all,linkcolor=blue,breaklinks=true,citecolor=blue,urlcolor=blue}
\usepackage[english]{babel}
\newcommand{\Sket}[1]{\bm{|}#1\bm{)}}  % requires \usepackage{bm}
\newcommand{\Sbra}[1]{\bm{(}#1\bm{|}}
\newcommand{\Sbraket}[2]{\bm{(}#1\bm{|}#2\bm{)}}

\newcommand{\brkt}[1]{\langle #1 \rangle }
\renewcommand{\braket}[2]{\langle #1 | #2 \rangle}

\newcommand{\one}{\mathds{1}}    % requires \usepackage{dsfonts}
\newcommand{\ones}{\mathcal{I}}
\renewcommand{\P}{\mathcal{P}}   % parity superoperator
 
\newcommand{\A}{\text{A}} % Andreev

\newcommand{\D}{\text{D}} % dot
\newcommand{\M}{\text{M}} % metal
\renewcommand{\S}{\text{S}} % superconductor
\newcommand{\T}{\text{T}} % tunneling term

\newcommand{\marker}[1]{\textcircled{\small{#1}}} % numbered figure markers\usepackage{float}

\begin{document}
\title{
  Transient transport spectroscopy of an interacting quantum dot
  \\
  proximized by a superconductor: Charge- and heat-currents after a switch
}
\author{Lara\,C.~Ortmanns}
\affiliation{
  Institute for Theory of Statistical Physics,
  RWTH Aachen, 52056 Aachen, Germany
}
\affiliation{
  JARA-FIT, 52056 Aachen, Germany
}
\affiliation{
  Department of Microtechnology and Nanoscience (MC2), Chalmers University of Technology, SE-41296 G\"oteborg
}

\author{Janine~Splettstoesser}
\affiliation{
  Department of Microtechnology and Nanoscience (MC2), Chalmers University of Technology, SE-41296 G\"oteborg
}

\author{Maarten\,R.~Wegewijs}
\affiliation{
	Institute for Theory of Statistical Physics,
	RWTH Aachen, 52056 Aachen, Germany
}
\affiliation{
	JARA-FIT, 52056 Aachen, Germany
}
\affiliation{
	Peter Gr{\"u}nberg Institut,
	Forschungszentrum J{\"u}lich, 52425 J{\"u}lich, Germany
}

%%%%%%%%%%%%%%%%%%%%%%%%%%%%%%%%%%%%%
\begin{abstract}
	We analyze the time-evolution of a quantum dot which is proximized by a large-gap superconductor
	and weakly probed using the \emph{charge and heat} currents into a wide-band metal electrode.
	We map out the full time dependence of these currents
	after initializing the system by a switch.
	We find that due to the proximity effect there are two simple yet distinct switching procedures
	which initialize a non-stationary mixture of the gate-voltage dependent eigenstates of the proximized quantum dot.
	We find in particular that the ensuing time-dependent heat current is a sensitive \emph{two-particle} probe
	of the interplay of strong Coulomb interaction and induced superconducting pairing.
	The pairing can lead to a suppression of charge and heat current decay
	which we analyze in detail.
	The analysis of the results makes crucial use of analytic formulas obtained using fermionic duality, a ``dissipative symmetry'' of master equations describing this class of open systems.
\end{abstract}
%%%%%%%%%%%%%%%%%%%%%%%%%%%%%%%%%%%%%

% LEAVE FOLLOWING EMPTY LINE HERE FOR mksubmit script !
\maketitle
%%%%%%%%%%%%%%%%%%%%%%%%%%%%%%%%%%%%%%%%%%%%%%%%%%%%%%%%%
\section{Introduction}
%%%%%%%%%%%%%%%%%%%%%%%%%%%%%%%%%%%%%%%%%%%%%%%%%%%%%%%%%

The dynamics of strongly interacting nanoelectronic structures continues to attract experimental~\cite{Gabelli2012Nov,Cottet2017Sep,Kleinherbers2022Feb} and theoretical~\cite{Andergassen2011May,Mitra2018Mar} interest.
The simplest type of nanostructure, in which time-dependent charge currents can be observed is a mesoscopic capacitor~\cite{Buttiker1993Jun,Buttiker1993Sep}, a confined quantum-dot like structure in contact with a single electronic reservoir
 driven out of equilibrium by a time-dependent gate voltage. Such nanostructures can serve as single-electron sources which are of high current interest~\cite{Feve2007May,Moskalets2011Sep,Pekola2013Oct,Bauerle2018Apr}. Time-dependent response of such systems is also of interest from a transport-spectroscopy perspective, in particular for strongly interacting quantum dots~\cite{Splettstoesser2010Apr,Andergassen2011May,Kennes2012Jun,Dittmann2018Apr,Stegmann2021Aug}, where not only transient charge currents but also transient heat currents~\cite{Schulenborg2016Feb,Ludovico2018Jan,Schulenborg2018Dec,Dashti2019Jul,Kamp2021Jan} reveal information that is not accessible from the stationary state transport.

%%%%%%%%%%%%%%%%%%%%%%%%%%%%%%%%%%%%%%%%%%%%%%%%%%%%%%%%%
\begin{figure}[h!]
	\centering
	\includegraphics[width=1.0\linewidth]{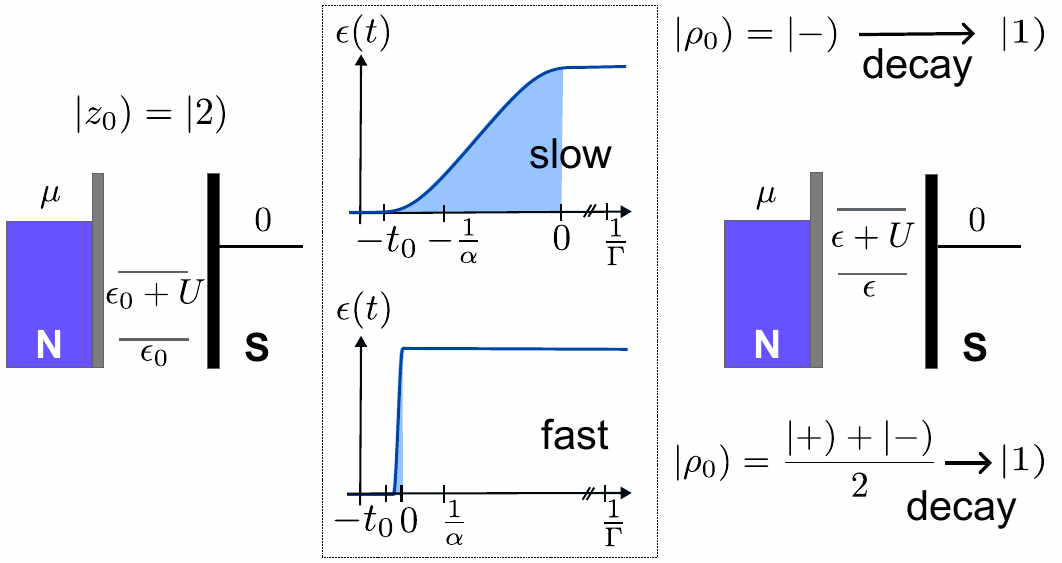}
	%\hfill
	\caption{Transient transport spectroscopy of a proximized quantum dot:
		For the example shown, an initialized state at $t=0$ decays to the final 1-electron $\Sket{1}$ state
		which is stationary for gate voltage $\epsilon$.
		The initialization starts at time $-t_0 < 0$ from a state which \emph{before the switch} is stationary at gate voltage $\epsilon_0$,
		in the example a doubly occupied state.
		It is transformed to a mixture of Andreev states $\rho_0$,
		either by a fast switch ($t_0 \ll \alpha^{-1}$, bottom inset) or a slow switch ($t_0 \gg \alpha^{-1}$, top inset).
		As explained in Sec.~\ref{sec:initialization}, 
		in both cases the \emph{effective} description by an initial mixture of Andreev energy eigenstates is appropriate. For the fast switch, this requires a charge and heat current readout that is coarse-grained on the time-scale $\alpha^{-1}$ of Cooper-pair oscillations, which is 
		much shorter than the dissipative transport time-scale (scale $\gamma_p^{-1}$) on which the metal probes the proximized dot.
	}%
	\label{fig:switch}%
\end{figure}
%%%%%%%%%%%%%%%%%%%%%%%%%%%%%%%%%%%%%%%%%%%%%%%%%%%%%%%%%

This can already be seen for the most simple realization of an interacting,  single-level quantum dot
coupled to a wide-band metal. When it is ‘’kicked” out of the stationary state, the
charge current decays with a characteristic time-scale denoted $\gamma_c^{-1}$ which is set by the microscopic electron tunneling rate $\Gamma$ but is modulated by all other parameters of the problem.
This time scale also appears in the heat current
which, as expected, features an additional time-scale $\gamma_p^{-1}$
because energy is a many-particle quantity due to electron-electron interaction on the dot. Remarkably, this scale depends \emph{only} on coupling strength $\Gamma$ but not on any other physical parameter~\cite{Schulenborg2016Feb}.

This is not an isolated observation but was noted in %a variety of 
different
contexts~\cite{Contreras-Pulido2012Feb,Saptsov2012Dec,Saptsov2014Jul,Schulenborg2014May}.
It was subsequently shown to be a consequence of \emph{fermionic duality}~\cite{Schulenborg2016Feb}
which is applicable to a large class of interacting dissipative nanostructures~\cite{Schulenborg2018Dec,Bruch21a}.
This intrinsically ``dissipative symmetry relation'' goes beyond mere time-scales:
it expresses a very strong restriction on the complete set of dynamical equations of an open system.
Importantly, it not only simplifies the calculation of the time-dependent solution,
but it also enables a deeper systematic analysis of the physical parameter dependence of this solution~\cite{Schulenborg2014May,Schulenborg2016Feb,Vanherck2017Mar,Schulenborg2017Dec,Schulenborg2018Dec,Schulenborg2020Jun,Bruch2021Sep,Monsel2022Feb}.
In particular, it allows one to analyze the \emph{amplitude} functions of the dynamics
which control the non-trivial competition of the various time-scales.
These amplitudes decide the actual time scale on which a system responds:
a slow decay term may well have negligible amplitude
and thus be irrelevant
depending on the gate voltages of the switching procedure and other control parameters.

It is an interesting question, how this complete picture of the decay of currents from the quantum dot  into a metal probe is altered when it is brought into contact with reservoirs with a more exotic structure.
In this paper we consider the case of an additional superconducting contact~\cite{Meden2019Feb}
which in general affects the quantum dot both through the quasiparticle tunnel rates and coherent pairing effects induced in the 0-2 charge sector of the quantum-dot Hilbert space.
This is highly relevant in view of the experimental advances in superconducting hybrid systems with quantum dots~\cite{DeFranceschi2010Oct,Lee2014Jan,Jellinggaard2016Aug,Hays2021Jul,Banerjee2022Jan} with controlled heat flow~\cite{Saira2007Jul,Martinez-Perez2015Apr}, fast gate control~\cite{Fatemi2021Dec,Hays2021Jul}, as well as time-resolved state readout~\cite{Hays2021Jul}. Superconducting contacts have also been exploited in combination with single-electron sources~\cite{Pekola2008Feb,vanZanten2016Apr,Dittmann2016Aug,Erdman2019Dec}.
Various ways of probing superconducting properties were theoretically addressed via short-time (finite-frequency) noise~\cite{Droste2015Mar,Braggio2011Jan} or higher-order cumulants of the short-time statistics~\cite{Rajabi2013Aug}
and waiting-time distributions~\cite{Wrzesniewski2020Apr}.
For weak coupling to the metal, the quantum dot exhibits an analog of Nambu-Goldstone and Higgs modes which
were recently studied, including their impact on heat transport~\cite{Kamp2019Jan}, decay time-scales and response to slow time-dependent driving~\cite{Kamp2021Jan}.
The latter response is also relevant for the analysis of charge pumping in superconducting quantum dot hybrids~\cite{Splettstoesser2007Jun,Hiltscher2011Oct,Moghaddam2012Mar},
for AC driving and spectroscopy~\cite{Siegl2023Mar,Ortega-Taberner2023Mar},
and for the adiabatic energy-response~\cite{Arrachea2018Jul}.
Also, the dynamics in the regime of strong coupling has been addressed, see Refs.~\cite{Wrzesniewski2021Apr,Taranko2021Apr}.
Finally, the dissipative spectral properties of proximized quantum dots also enter into their open-system topological properties of full counting statistics~\cite{Herrig2020Dec}.

Clearly, for applications of such superconducting hybrid structures it is of central importance
to have a complete understanding of \emph{both time-scales and amplitudes} of the transport dynamics after a basic switch illustrated in Fig.~\ref{fig:switch}.  This is the scope of the present paper.
Our analysis focuses on the conceptually simplest case
of a single-level quantum dot proximized by a superconductor with a large gap (no quasiparticle transport)
probed by a weakly coupled wide-band metal (only single-electron processes).
Interestingly, we find that already in this simple case, this switching procedure itself requires special attention,
since in a proximized dot, the energy eigenstates of the proximized structure
are gate-voltage dependent.
We consider two distinct physical preparation procedures,
denoted ``fast switch'' and ``slow switch'', respectively,
and map out their full time-evolution of charge and heat currents
based on master and transport equations derived earlier~\cite{Sothmann2010Sep}.
We make use of the analytical solution of these equations reported in a recent work~\cite{Ortmanns22a}
which did not address the specific initialization.
This work fully exploited the above mentioned fermionic duality which is applicable to \emph{proximized} quantum dot systems~\cite{Schulenborg2016Feb,Schulenborg2018Dec}
finding that the interesting interplay of strong repulsive interaction and strong induced superconducting pairing is exhibited \emph{only} in the time-scale characteristic of the quasiparticle charge decay ($\gamma_c^{-1}$):
similar to systems without a superconductor,
the Andreev-state parity decay time ($\gamma_p^{-1}$) is insensitive to any parameter except the metal contact tunnel-rate constant $\Gamma$.
In particular the proximity of the large-gap superconductor is of no effect for this rate
as also noted in Ref.~\cite{Kamp2021Jan}.

The present complete analysis of this solution starting from Ref.~\cite{Ortmanns22a} is, however, mostly concerned with the decisive parameter dependence of the decay \emph{amplitudes}
and thereby extends the \emph{transient} charge and heat \emph{transport spectroscopy} in response to electrostatic switching
presented in Ref.~\cite{Schulenborg2016Feb} to Andreev states.
This dependence becomes quite nontrivial due to superconducting pairing:
the amplitudes of the transient charge and heat decay exhibit a variety of features
which strongly depend on the five different energy scales of the problem
and on whether the fast or slow switching procedure is used.

The paper is organized as follows. After reviewing the description of the proximized quantum dot system in Sec.~\ref{sec:model}, we analyze in Sec.~\ref{sec:initialization} two ideal experimental initialization schemes using gate-voltage switching.
In Sec.~\ref{sec:solution} we review the transport equations of Ref.~\cite{Sothmann2010Sep} and
analyze their solution reported in Ref.~\cite{Ortmanns22a}
in terms of fermionic-duality dictated variables.
We then systematically discuss the time-dependent spectroscopy in Secs.~\ref{sec:spectroscopy} and explain a number of salient features in Sec.~\ref{sec:discussion} before concluding in Sec.~\ref{sec:summary}.
Throughout the paper we use units such that $k_\text{B}=\hbar = |e|= 1$.

% LEAVE FOLLOWING EMPTY LINE HERE FOR mksubmit script !
%%%%%%%%%%%%%%%%%%%%%%%%%%%%%%%%%%%%%%%%%%%%%%%%%%%%%%%%%
\section{Proximized quantum dot\label{sec:model}}
%%%%%%%%%%%%%%%%%%%%%%%%%%%%%%%%%%%%%%%%%%%%%%%%%%%%%%%%%

Following Ref.~\cite{Sothmann2010Sep,Ortmanns22a} we describe the system
sketched in Fig.~\ref{fig:switch}
using a quantum-dot Hamiltonian
$
	H_\D' = \epsilon N + U N_{\uparrow} N_{\downarrow}
$,
with a single level $\epsilon$ controlled by a gate voltage and Coulomb repulsion $U \geq 0$.
Here $N=N_{\uparrow}+N_{\downarrow}$ is the electron number operator with $N_{\sigma}=d^\dagger_\sigma d_\sigma$.
The dot is tunnel coupled by
$
	H_\T = \sum_{k\sigma} \sqrt{\Gamma/(4\pi)}(d_\sigma^\dag c_{k\sigma}+\text{h.c.})
$
to a flat-band metallic reservoir $H_\M=\sum_{k\sigma} \omega_k c_{k\sigma}^\dag c_{k\sigma}$
which is held at temperature $T$ and electrochemical potential $\mu$.
The electron number operator on the metal is denoted $N_\M=\sum_{k\sigma} c_{k\sigma}^\dag c_{k\sigma}$.
The quantum dot is additionally coupled to a superconductor held at zero electrochemical potential, $\mu_\S=0$.
In the limit of large superconducting gap exceeding coupling and temperature $\Delta \gg T, \Gamma$
its influence on the dot is described~\cite{Governale2008Apr,Affleck2000Jul,Tanaka2007May,Meng2009Jun,Karrasch2011Oct}
by a Hamiltonian pairing term with real-valued pairing amplitude $\alpha$
\begin{align}
	H_\S = - \tfrac{1}{2} \alpha d^\dag_\uparrow d_\downarrow^\dag + \text{h.c}
	\ .
	\label{eq:HS}%
\end{align}%
This effective $\Delta \to \infty$ model neglects quasiparticles in the superconductor.
Here and in Sec.~\ref{sec:solution} it is important to allow $\alpha$ to have an arbitrary sign
but in Secs.~\ref{sec:initialization}, \ref{sec:spectroscopy} and \ref{sec:discussion} we take $\alpha$ positive as usual.

\begin{figure}[t]%[tb]
	\includegraphics[width=1.0\linewidth]{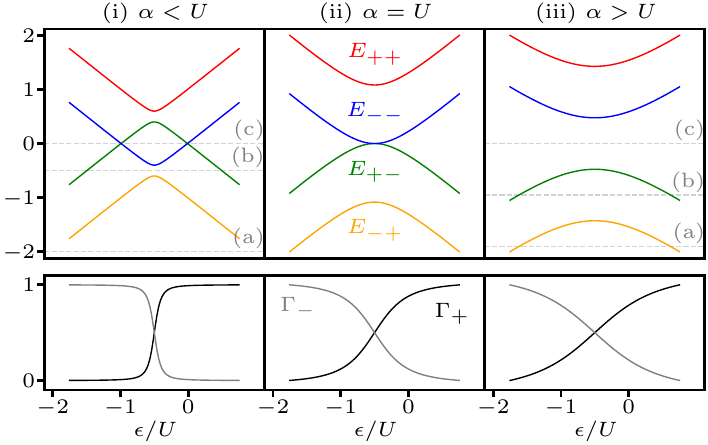}
	\caption{
		Energy thresholds $E_{\eta,\tau}/U$ for Andreev transitions
		and effective rates $\Gamma_\tau/\Gamma$,
		as function of gate voltage $\epsilon$ relative to $\mu_\S=0$ of the superconductor.
		Qualitatively different biasing conditions (a)-(c) are indicated by dashed horizontal lines (grey).
		For weak pairing $\alpha < U$
		the levels $E_{+-}$ and $E_{--}$ at $\delta=0$ are split by $U-\alpha$,
		but they cross at $|\delta|=U$,
		reversing the splitting. For strong pairing $|\alpha| \geq U$, we always have $E_{--}>E_{+-}$ without crossings and a minimum splitting of $\alpha-U$ for any gate voltage.
	Parameters:  (i) $\alpha/U=0.2$  (ii) $1.0$ and (iii) $2.0$.
	}
	\label{fig:thresholds}
\end{figure}

The coupling of the dot to the metal probe is assumed to be weak,
both relative to the metal thermalization, $\Gamma \ll T$,
and to the gap induced on the proximized dot, $\Gamma \ll |\alpha|$,
as for example in a recent relevant experiment~\cite{Jellinggaard2016Aug}.
We are interested in the charge current
$
I_N=\partial_t \langle N_\M\rangle
$
(electron particle current)
and the heat current
$
I_Q=\partial_t \langle H_\M-\mu N_\M\rangle
$,
flowing by convention into the metal.
Our goal is to understand how these two quantities observable in the metal probe time-dependently probe the dot-superconductor system.

In the Hamiltonian describing this \emph{proximized dot},
\begin{align}
	H_\D = H_\D'+H_\S
	= \sum_{\tau} E_\tau \ket{\tau}\bra{\tau} + E_1 \sum_\sigma \ket{\sigma}\bra{\sigma}
	\label{eq:H_D}
\end{align} 
the pairing generates an Andreev splitting $\delta_\A \geq |\alpha|$
of the discrete 0- and 2-electron levels~\cite{Rozhkov2000Sep}.
For $\alpha \neq 0$ this splitting differs from the detuning energy $\delta$ as follows:
\begin{align}
	E_\tau & = \frac{1}{2} \big( \delta + \tau \delta_\A \big)
	,\quad
	\begin{cases}
		\delta   &= 2\epsilon+U
		\\
		\delta_\A  &= \sqrt{\delta^2 + \alpha^2}
	\end{cases}
	.
	\label{eq:pm-energies}
\end{align}%
The corresponding states are hybridized into Andreev states with isospin label $\tau = \pm$:
\begin{align}
	\ket{\tau}
	&=
	\sqrt{\tfrac{1}{2}\Big[ 1-\tau \frac{\delta}{\delta_\A} \Big] } \ket{0}
	-\text{sign}(\alpha) \, \tau
	\sqrt{\tfrac{1}{2}\Big[ 1+\tau \frac{\delta}{\delta_\A} \Big] } \ket{2}
	\notag
	\\
	& = \sqrt{\Gamma_{-\tau}/\Gamma } \ket{0}
	-\text{sign}(\alpha) \, \tau
	\sqrt{\Gamma_{ \tau}/\Gamma } \ket{2}	
	.
	\label{eq:pm_states}
\end{align}
This hybridization is maximal around the particle-hole symmetry point of the dot, $\epsilon=-U/2$,
where $\delta=0$ such that the splitting is minimal, $\delta_\A = |\alpha|$.
Unaffected by pairing are the 1-electron states of the dot
--uniform mixtures of spin states $\ket{\sigma}$-- with energy
\begin{align}
	E_1=\epsilon=\tfrac{1}{2}\big(\delta-U\big)
	\label{eq:E_1}
	.
\end{align}

\begin{table}
	\caption{
		Energy thresholds ($\gtrless$) for transitions ($\rightleftarrows$)
		by transfer of  $^e_h$ ($^h_e$) from the metal to the dot.
	}
	\begin{ruledtabular}
		\begin{tabular}{ c c c c }
			$\eta\tau$ & Threshold  & Rate  & Transition $\overunderset{e}{h}{\rightleftarrows}$
			\\
			\hline
			$++$ & $\mu \gtrless E_{++}$& $\Gamma_+$ & $1 \rightleftarrows + $
			\\[1ex]
			$--$ & $\mu \gtrless E_{--}$& $\Gamma_+$ & $- \rightleftarrows 1 $ 
			\\[1ex]
			$+-$ & $\mu \gtrless E_{+-}$& $\Gamma_-$ & $1 \rightleftarrows - $ 
			\\[1ex]
			$-+$ & $\mu \gtrless E_{-+}$& $\Gamma_-$ & $+ \rightleftarrows 1 $ 
			%\\[1ex]
			%\hline \hline
		\end{tabular}
	\end{ruledtabular}
	\label{table:addition_energies}
\end{table}

The charge and heat transport into the metal probe are expected to change in two situations.
First, the probabilities featuring in the Andreev-states~\eqref{eq:pm_states} define effective rates $\Gamma_\tau$ 
\begin{equation}
	\frac{\Gamma_{\tau}}{\Gamma}
	=
	\frac{1}{2} \Big( 1 + \tau \frac{\delta}{\delta_\A} \Big)
	.
	\label{eq:Gamma_pm}
\end{equation}
for tunneling into state $\tau=\pm$,
which are plotted in Fig.~\ref{fig:thresholds} as function of gate voltage
entering $\epsilon$ through $\delta$ [Eq.~\eqref{eq:E_1}].
Both rates show resonant behavior at $|\delta| = 0$
broadened by the pairing energy $|\alpha|$.
In this way the hybridized dot model
incorporates~\cite{Pala2007Aug,Governale2008Apr,Sothmann2010Dec,Droste2015Mar} the \emph{superconductor resonance}
at which electron pairs are exchanged with the quantum dot.

Second, whenever the metal electrochemical potential $\mu$ matches one of the four Andreev-transition energies
\begin{align}
	E_{\eta, \tau}
	= \eta \Big( E_\tau - E_1 \Big)
	= \tfrac{1}{2} \Big( \eta \tau \delta_\A +  \eta U \Big)
	.
	\label{eq:energies_transition}
\end{align}%
there is \emph{resonance} with the \emph{metal} broadened by its thermal energy $T$.
Here $\tau = \pm$ indicates the involved Andreev state and $\eta = \pm$ the direction $1 \rightleftarrows \tau$ of the corresponding state transition:
The dot transition $1 \rightleftarrows \tau$ can be realized by an electron transfer from the metal to the dot, when $\mu \geq E_{\eta, \tau}$.
Importantly, due to the pairing $\alpha$, each state transition $1 \rightleftarrows \tau$ can \emph{also} be realized by an electron transfer in the \emph{opposite} direction, from the dot to the metal, when $\mu \leq E_{\bar{\eta},\tau}$.
Here and below the overbar indicates the opposite value, $\bar{\eta}=-\eta$.
These transitions, their corresponding particle transfers and energy thresholds are listed Table \ref{table:addition_energies}.
The dependence of the energy thresholds on gate voltage is also plotted in Fig.~\ref{fig:thresholds} and qualitatively different biasing conditions studied later on are indicated. Note that the values for $\mu$ that qualify as small, intermediate or large biases depends on the value of $\alpha$ as becomes clear from Fig.~\ref{fig:thresholds}.

% LEAVE FOLLOWING EMPTY LINE HERE FOR mksubmit script !
%%%%%%%%%%%%%%%%%%%%%%%%%%%%%%%%%%%%%%%%%%%%%%%%%%%%%%%%%
\section{State initialization\label{sec:initialization}}
%%%%%%%%%%%%%%%%%%%%%%%%%%%%%%%%%%%%%%%%%%%%%%%%%%%%%%%%%

We analyze the time-dependent transport resulting from
an initial mixture of energy eigenstates of $H_\D$, described by a density operator denoted $\rho_0$.
We now discuss how such a mixture is physically prepared by controlling experimental parameters.
This issue was left open in Refs.~\cite{Sothmann2010Sep,Ortmanns22a} where an \emph{arbitrary} unspecified mixture $\rho_0$ of energy states was considered.

%%%%%%%%%%%%%%%%%%%%%%%%%%%%%%%%%%%%%%%%%%%%%%%%%%%%%%%%%
\subsection{Gate switch and pairing}\label{sec:switch}
%%%%%%%%%%%%%%%%%%%%%%%%%%%%%%%%%%%%%%%%%%%%%%%%%%%%%%%%%
We focus on the experimental situation where the initial state, $\rho_0$,  is prepared by switching a gate voltage.
We therefore consider possible extensions of the simple scenario analyzed in Ref.~\cite{Schulenborg2016Feb}
where an interacting quantum dot was probed by a metal after a sudden change of the dot level $\epsilon_0 \to \epsilon$
as sketched in Fig.~\ref{fig:switch}.
It is assumed that before this switch the dot has decayed to its stationary mixed state  at $\epsilon_0$ denoted $z_0$.
After switching to $\epsilon$ the dot decays to its new stationary mixed state denoted by $z$.
Throughout the paper we will indicate the value of quantities such as $z_0$
before the switch, namely at $\epsilon_0$, by a subscript $0$
as opposed to its value after the switch, the stationary state $z$ at $\epsilon$.
In addition, also the initial state prepared by the switch, $\rho_0$, that we discuss here, is indicated by a subscript $0$.

\begin{figure}[tb]
	\centering
	\includegraphics[width=0.8\linewidth]{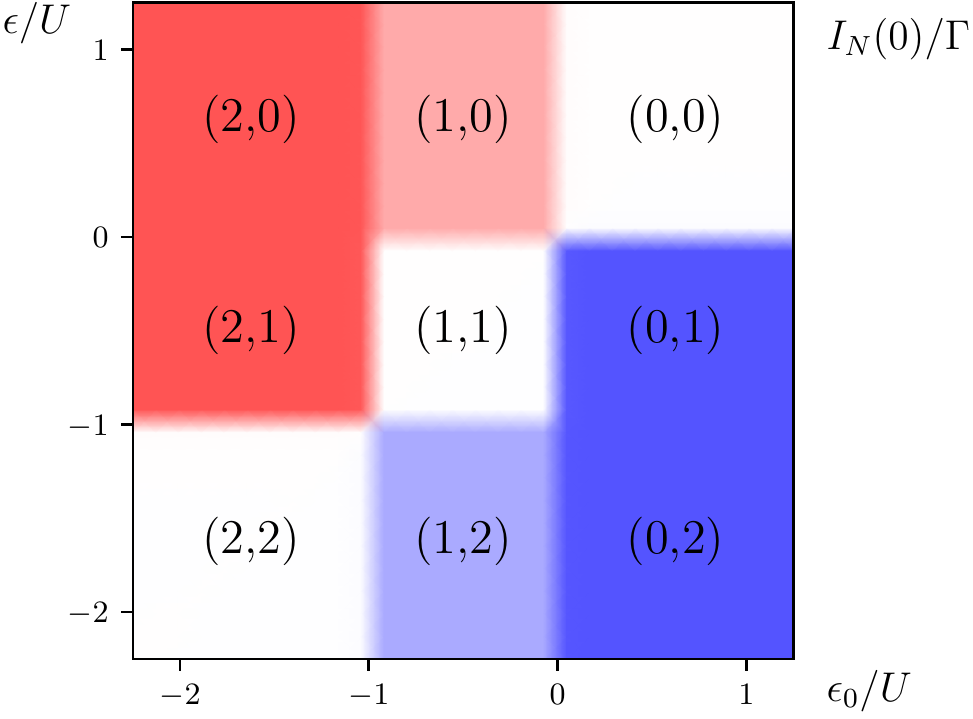}
	\caption[b]{
		Transient charge transport spectroscopy (no superconductor, $\alpha=0$).
		Plotted is the transient charge current $I_N(0)$ 
		flowing right after a gate-voltage switch $\epsilon_0 \to \epsilon$ in units of the tunnel rate $\Gamma$.
		For low $T \ll U$ this results in a decay of the dot charge
		from a nearly integer initial stationary value $\brkt{N}_{z_0}$
		towards another nearly integer stationary value $\brkt{N}_z$.
		Possible values of $(\brkt{N}_{z_0}, \brkt{N}_z )$ are indicated. Red indicates positive, blue negative currents.
	}%
	\label{fig:configurations}%
\end{figure}

As a result of this simple procedure, in the \emph{absence} of the superconductor ($\alpha=0$),
the predictions for the charge current decay (chosen here as one example for a transient observable) 
can be analyzed using a \emph{single} transient transport spectroscopy plot for all possible gate voltage switches,
see Fig.~\ref{fig:configurations}.
This conveniently represents the results for the transient charge current amplitude for all time-dependent switching scenarios (gate-voltage pairs)
and will be reviewed in more detail in Sec.~\ref{sec:spectroscopy}.
The basic idea is simple:
Up to thermal smearing 9 basic switches are possible
since depending on $\epsilon_0$ resp. $\epsilon$,
the stationary dot state has
$\brkt{N}_{z_0}$ resp. $\brkt{N}_z$
$=0,1$ or $2$ electrons.
As seen in Fig.~\ref{fig:configurations},
a (positive/negative) transient charge current ensues (red/blue) when switching between regimes with
$\brkt{N}_{z_0} \neq \brkt{N}_z$.
In the regimes $\brkt{N}_{z_0} = \brkt{N}_z$
there is no transient charge current.

%%%%%%%%%%%%%%%%%%%%%%%%%%%%%%%%%%%%%%%%%%%%%%%%%%%%%%%%%
%%%%%%%%%%%%%%%%%%%%%%%%%%%%%%%%%%%%%%%%%%%%%%%%%%%%%%%%%

Interestingly, the extension of this scenario to include the large-gap superconductor is not unique.
The reason is that the Andreev states~\eqref{eq:pm_states} 
for the gate voltage before and after the switch differ:
In contrast to Eq.~\eqref{eq:pm_states} at $\epsilon$,
we have at $\epsilon_0$
\begin{align}
	\ket{\tau_0} =
	\sqrt{\tfrac{1}{2}\Big[ 1-\tau \frac{\delta_0}{\delta_{\A0}} \Big] } \ket{0}
	-\text{sign}(\alpha) \tau
	\sqrt{\tfrac{1}{2}\Big[ 1+\tau \frac{\delta_0}{\delta_{\A0}} \Big] } \ket{2}
	\label{eq:pm_states0}
\end{align}
denoting $\delta_0 = 2 \epsilon_0+U$ and $\delta_{\A0}=\sqrt{\delta_0^2+\alpha^2}$.
In Sec.~\ref{sec:slow_switch} and \ref{sec:fast_switch} we discuss in detail two distinct, experimentally relevant
ways of preparing initial states by gate switching, resulting in different transient transport.

In contrast to the no-superconductor case~\cite{Schulenborg2016Feb}
these two cases also differ by the time scale for probing.
Indeed, it is important to note that this requires careful discussion:
we consider switch scenarios for which
the simple description by occupations alone as in Refs.~\cite{Sothmann2010Sep,Ortmanns22a}
cannot be applied without further consideration of what aspect of the probe currents one aims to describe.
For example, if one switches $\epsilon_0\to\epsilon$
on a timescale comparable to or smaller than the inverse induced pairing gap $\alpha^{-1}$ and at the same time aims at a readout at time scales of the order of $\alpha^{-1}$, coherent superpositions between even-parity states would be relevant for the system state dynamics after a gate switch, see Refs.~\cite{Kamp2019Jan,Kamp2021Jan}.
In this case, a state $\ket{\tau_0}$ prepared as the stationary state at gate voltage $\epsilon_0$ before the switch is a superposition --not a mixture-- of energy states $\ket{\tau}$ for the new gate voltage $\epsilon$ after the switch.
This happens with maximal amplitudes if, as in Fig.~\ref{fig:switch},
we start far from resonance ($|\delta_0| \gg \alpha$),
meaning that the stationary state before the switch is either
$\ket{-_0} \approx\ket{0}$
or 
$\ket{-_0} \approx \ket{2}$,
and then rapidly switch the gate voltage to resonance ($|\delta| \ll \alpha$)
where the charge states are uniform superpositions $\ket{0,2}=(\ket{+}\pm \ket{-})/\sqrt{2}$.
(Note that up to a global phase,
$\ket{2}   =	\sum_{\tau}\tau
\sqrt{\tfrac{1}{2}[ 1 + \tau {\delta}/{\delta_\A} ] } \ket{\tau}$
and $
\ket{0}   = \sum_{\tau}
\sqrt{\tfrac{1}{2}[ 1 - \tau {\delta}/{\delta_\A} ] } \ket{\tau}
$.)
In the present paper, we focus instead on two experimentally relevant scenarios for which the simple description of~\cite{Sothmann2010Sep,Ortmanns22a} remains applicable.

%%%%%%%%%%%%%%%%%%%%%%%%%%%%%%%%%%%%%%%%%%%%%%%%%%%%%%%%%
\subsection{Slow switch}\label{sec:slow_switch}
%%%%%%%%%%%%%%%%%%%%%%%%%%%%%%%%%%%%%%%%%%%%%%%%%%%%%%%%%

We first consider a situation denoted as ``slow switch''
where the switching time of the gate voltage is much longer than the pairing timescale $\alpha^{-1}$,
but much shorter than the time scale of dissipative tunneling $\Gamma^{-1}$ to the probe metal. In a realistic setting~\cite{Jellinggaard2016Aug} with $\Gamma\sim\mu$eV and $\alpha\sim$meV, this would correspond to switching times of tens of fs. Alternatively, it could be realized by  temporarily decoupling the metal $\Gamma \to 0$
to prevent dissipative effects during the switching procedure, lifting the upper limit on the switching time.
During a gate-voltage switch on a time scale much larger than $\alpha^{-1}$ 
the states $\ket{\tau_0}$ will evolve unitarily to $\ket{\tau}$:
\begin{align}
	\ket{\tau_0} \to \ket{\tau} = U(\epsilon,\epsilon_0) \ket{\tau_0}
	\label{eq:tau-unitary}
\end{align}
For the example at the end of the previous section,
where the state before the switch is a pure even-parity state,
this implies that one prepares a \emph{pure} energy state
$\ket{\tau_0}\bra{\tau_0} \to \ket{\tau}\bra{\tau}
=(\ket{0}-\tau \ket{2})(\bra{0}-\tau \bra{2})/2$ 
for $\tau=\pm$
as illustrated in Fig.~\ref{fig:switch} (top panel)
instead of a complete mixture (as for the fast switch in the following section).
Including the odd-parity part into our discussion,
this slow switch maintains the stationary mixing coefficients
but alters the basis vectors:
\begin{align}%
	z_0 &=
	\sum_{\tau=\pm}
	z_{0\tau} \ket{\tau_0}\bra{\tau_0} + z_{01} \tfrac{1}{2}\sum_\sigma \ket{\sigma}\bra{\sigma}
	\label{eq:rho0_slow}%
	\\
	\to \quad
	\rho_0 &=
	\sum_{\tau=\pm}
	z_{0\tau} \ket{\tau}\bra{\tau}+ z_{01} \tfrac{1}{2}\sum_\sigma \ket{\sigma}\bra{\sigma}
	\notag
	.	
\end{align}%
With $\rho_0$, we indicate the initial state of the decay dynamics, in which the system is found at the end of the switch.
Since the slow-switch procedure generates no off-diagonal elements in the new, final energy basis,
we can compute the ensuing evolution and transport using the equations of Ref.~\cite{Ortmanns22a}.

%%%%%%%%%%%%%%%%%%%%%%%%%%%%%%%%%%%%%%%%%%%%%%%%%%%%%%%%%
\subsection{Fast switch}\label{sec:fast_switch}
%%%%%%%%%%%%%%%%%%%%%%%%%%%%%%%%%%%%%%%%%%%%%%%%%%%%%%%%%

The second situation we consider is denoted ``fast switch''
where the gate voltage is changed on a timescale much shorter than $\alpha^{-1}$.
In this case we consider a coarse-grained time-dependent readout 
in which features occurring on the time-scale $\alpha^{-1}$ are averaged out.
For example, in Ref.~\cite{Hays2021Jul} an experimental readout of the time-evolution of Andreev-state occupations was performed at a resolution of down to few nanoseconds.

If we instantly switch the gate voltage from $\epsilon_0 \to \epsilon$, the initial state is in general a superposition of the proximized energy eigenstates at the new gate voltage $\epsilon$:
\begin{align}
	\ket{\tau'_0} = \sum_{\tau} \ket{\tau} \braket{\tau}{\tau_0'}\ .
\end{align}
Their similarity is quantified by transition probabilities
\begin{align}
	|\braket{\tau}{\tau'_0} |^2 = \tfrac{1}{2} \big[ 1 + \tau \tau' \theta \big]
	,\quad \tau,\tau'=\pm 
\end{align}
which can be expressed in terms of a single \emph{probability-bias} parameter $\theta \in [-1,1]$.
Inserting Eqs.~\eqref{eq:pm_states} and~\eqref{eq:pm_states0}:
\begin{align}
	\theta
	= \tfrac{1}{2} \sum_{\tau\tau'} \tau \tau' |\braket{\tau}{\tau'_0} |^2
	=\frac{ \delta \delta_0 + \alpha^2 }{\delta_\A \delta_{\A 0}}
	.
	\label{eq:theta}
\end{align}
When we subsequently let the system evolve to the new stationary situation,
the state and the currents probing this evolution will decay on a timescale $\Gamma^{-1}$ while oscillating on the much shorter pairing-induced timescale $\alpha^{-1}$ since we assume $\alpha \gg \Gamma$ [Sec.~\ref{sec:model}].
For this switching procedure we consider results of current measurements in the metal probe which are \emph{coarse-grained}, namely time-averaged over these rapid oscillations.

To describe only this longitudinal decay needed for the coarse-grained transient currents
we can modify the stationary state prepared before the switch
by keeping only the part that is energy-diagonal in the energy basis $\ket{\tau}$ \emph{after} the switch:
\begin{align}
	\ket{\tau'_0}\bra{\tau'_0}
	\to
	\sum_{\tau} | \braket{\tau}{\tau_0'} |^2 \cdot
	\ket{\tau}\bra{\tau}
\end{align}
This coarse-grained description in time
corresponds to fast complete decoherence in the final energy basis.
In a Bloch vector picture this can be understood [Eq.~\eqref{eq:bloch} and Eq.~\eqref{eq:rho0_fast_A}] as ignoring the rapid precession of the transverse Bloch components (off-diagonal elements) which occurs during the slow relaxation of the longitudinal Bloch component (diagonal elements).
Note that this does not mean that effects of superconducting coherence
are lost (which would be decoherence in charge basis).

Returning again to the example at the end of Sec.~\ref{sec:switch}, this fast switch procedure implies
that if we switch to resonance ($|\delta| \ll \alpha$)
coming from afar ($|\delta| \gg \alpha$)
we consider the preparation of a \emph{completely mixed state}:
$\ket{\tau'_0}\bra{\tau'_0} \to \tfrac{1}{2} \sum_\tau \ket{\tau}\bra{\tau}$
as was illustrated in Fig.~\ref{fig:switch} (bottom part).
Including the odd-parity part into our discussion,
the fast switch maps the stationary probabilistic mixture in the $\ket{\tau_0}$ basis
before the switch to a mixture in the basis after the switch
\begin{align}%
	z_0 &=
	\sum_{\tau'}
	z_{0\tau'} \ket{\tau_0'}\bra{\tau_0'} + z_{01} \tfrac{1}{2} \sum_\sigma \ket{\sigma}\bra{\sigma}
	\label{eq:rho0_fast}%
	\\
	\to
	%\quad
	\rho_0 &=
	\sum_{\tau}
	\Big( \sum_{\tau'} | \braket{\tau}{\tau_0'} |^2 z_{0\tau'} \Big)
	\ket{\tau}\bra{\tau}
	 + z_{01} \tfrac{1}{2} \sum_\sigma \ket{\sigma}\bra{\sigma}
\notag
\end{align}%
The evolution of \emph{these} initial occupations using the equations of Ref.~\cite{Ortmanns22a}
gives the \emph{coarse-grained} transport current after a fast switch $\epsilon_0 \to \epsilon$.

%%%%%%%%%%%%%%%%%%%%%%%%%%%%%%%%%%%%%%%%%%%%%%%%%%%%%%%%%
\subsection{Time-dependent spectroscopy and pairing}
%%%%%%%%%%%%%%%%%%%%%%%%%%%%%%%%%%%%%%%%%%%%%%%%%%%%%%%%%

Switches \eqref{eq:rho0_slow} and \eqref{eq:rho0_fast} describe the initialization of two distinct experiments
to which the description of Ref.~\cite{Sothmann2010Sep} applies.
Given a fixed $\alpha$,
each procedure is completely characterized by the pair of gate voltages $(\epsilon_0,\epsilon)$
 by giving the parameter $\theta$ [Eq.~\eqref{eq:theta}].
This allows to extend the complete analysis of the dynamics
in terms of two-dimensional spectroscopy diagrams~\cite{Schulenborg2016Feb} as in Fig.~\ref{fig:switch}
 to the case involving a superconductor.
A systematic comparison of the results is warranted
since for the fast switch the induced pairing $\alpha$ affects \emph{both}
the final stationary mixed state ($z$) of the transient dynamics
\emph{and} the initial state $\rho_0$ [prepared from the initial stationary mixed state $z_0$ by Eq.~\eqref{eq:rho0_fast}].
By contrast, for the slow switch we lose the dependence on overlap of initial and final energy eigenstates:
Indeed, if one formally sets $\theta=1$ independent of $\alpha,\epsilon_0,\epsilon$
such that $|\braket{\tau}{\tau'_0} |^2 = \delta_{\tau,\tau'}$
then Eq.~\eqref{eq:rho0_fast} reduces to Eq.~\eqref{eq:rho0_slow}.
This dependence on the \emph{states} is of key interest:
Without the superconductor ($\alpha=0$) at low temperature
the states $z$ and $\rho_0$ are at best a mixture of \emph{one} even-parity and the odd-parity state. When we include a superconductor and tune it to its resonance, the pairing
creates a ``shortcut'' in the decay sequence by connecting \emph{two even-parity} states.
This effect of the superconducting coherence of the energy states in the initial energy mixtures
can effectively \emph{counteract} the time-dependent decay into the metal as we will see.

% LEAVE FOLLOWING EMPTY LINE HERE FOR mksubmit script !
%%%%%%%%%%%%%%%%%%%%%%%%%%%%%%%%%%%%%%%%%%%%%%%%%%%%%%%%
\section{Solution of transport equations}\label{sec:solution}
%%%%%%%%%%%%%%%%%%%%%%%%%%%%%%%%%%%%%%%%%%%%%%%%%%%%%%%%

%%%%%%%%%%%%%%%%%%%%%%%%%%%%%%%%%%%%%%%%%%%%%%%%%%%%%%%%
\subsection{Master equation and current formulas}
%%%%%%%%%%%%%%%%%%%%%%%%%%%%%%%%%%%%%%%%%%%%%%%%%%%%%%%%
Following Ref.~\cite{Ortmanns22a}
we first review the transport equations of Ref.~\cite{Sothmann2010Sep}
describing the setup in Fig.~\ref{fig:switch}.
The time evolution of an initial state $\rho(0)=\rho_0$ of the dot
which is diagonal in the energy basis follows the rate equations
\begin{subequations}%
	\begin{align}%
		\frac{d\rho_\tau}{dt} & = W_{\tau 1} \rho_1 - W_{1\tau} \rho_{\tau},
		\label{_tau}                                                    \\
		\frac{d\rho_1}{dt}    & = \sum_{\tau} W_{1 \tau} \rho_{\tau} - 
		\sum_{\tau} W_{\tau 1} \rho_1							\ .	
	\end{align}%
	\label{eq:master_equation}%
\end{subequations}%
Here, the time-dependent state of the proximized dot is denoted $\rho(t)$
and its diagonal elements
$\rho_\tau(t)$ with $\tau=\pm$ are the occupation probabilities of even-parity states $\ket{\tau}$
after the switch
and $\rho_1(t)$ is the occupation of the odd-parity state
(a mixture of spin $\sigma=\uparrow,\downarrow$).
The charge and energy current flowing into the metal probe read
\begin{align}
	I_N & = \sum_{\eta\tau} \eta \big[ W_{1,\tau}^\eta \rho_\tau + W_{\tau,1}^\eta \rho_1 \big]
	\label{eq:I_N_energy_basis}
	,
	\\
	I_E & =
	\sum_{\eta\tau}
	\big[
	(E_\tau-E_1) W_{1,\tau}^\eta \rho_\tau
	+
	(E_1-E_\tau) W_{\tau,1}^\eta \rho_1
	\big]
	\label{eq:I_E_energy_basis}
	,
\end{align}
and the heat current reads $I_Q=I_E-\mu I_N$.
The transition rates are
$
W_{1,\tau} = \sum_\eta W_{1,\tau}^\eta
$
and
$
W_{\tau,1} = \sum_\eta W_{\tau,1}^\eta
$
whereas the transport rates
\begin{subequations}
	\begin{align}
		W_{1,\tau}^\eta & =\phantom{\tfrac{1}{2}}
		\Gamma_{\eta\tau}       f^{-\eta}       (E_{\eta,\tau}-\mu)
		,
		\\
		W_{\tau,1}^\eta & =\tfrac{1}{2}
		\Gamma_{\bar{\eta}\tau} f^{+\bar{\eta}} (E_{\bar{\eta},\tau}-\mu)
		,
	\end{align}%
	\label{eq:rates_transport}%
\end{subequations}%
keep track of whether an electron,
present in the metal
with probability $f^{-\eta}(x) = (e^{-\eta x/T}+1)^{-1}$,
is transferred to ($\eta=-$) or from ($\eta=+$) the proximized dot
from (to) the metal, 
where
$\bar{\eta} \equiv - \eta$ as before.
These rates depend on $E_{\eta, \tau}=\tfrac{1}{2} ( \eta \tau \delta_\A +  \eta U )$, transition energies~\eqref{eq:energies_transition}, and on $\Gamma_\tau$, the effective, gate-voltage dependent rate~\eqref{eq:Gamma_pm} for tunneling into energy eigenstate $\tau=\pm$,
see Table~\ref{table:addition_energies}.

%%%%%%%%%%%%%%%%%%%%%%%%%%%%%%%%%%%%%%%%%%%%%%%%%%%%%%%%
\subsection{Solution by fermionic-duality invariance}\label{sec:solution_duality}
%%%%%%%%%%%%%%%%%%%%%%%%%%%%%%%%%%%%%%%%%%%%%%%%%%%%%%%%

The evolution and transport equations can be solved standardly
by expanding the rate superoperator $W$ in its three left and right eigenvectors
$W=- \sum_{x}\gamma_x \Sket{x}\Sbra{x'}$
where $x$ labels $p$ (parity), $c$ (charge) and $z$ (zero, $\gamma_z\equiv 0$).
The prime distinguishes (operators corresponding to the) left and right eigenvectors for the same eigenvalue,
$z \neq z'$, $c\neq c'$, and $p \neq p'$.
Computing the eigenvectors and the solution in this way
leaves unexploited a strong restriction that applies to this class of problems~\cite{Schulenborg2016Feb}:
The rate superoperator obeys \emph{fermionic duality}
	\begin{equation}
		W + \tfrac{1}{2} \gamma_p \, \ones = - \big[ \P \overline{ \big( W + \tfrac{1}{2}\gamma_p \, \ones \big) } \P \big]^\dag
	\end{equation}
with $\gamma_p = \Gamma$.
When instead making use of this property as in Ref.~\cite{Ortmanns22a}
one obtains a much simpler and more insightful form of the time-dependent solution which we review below.
For the purpose of the present paper it is only relevant to note that the duality mapping of superoperators like $W$
involves a mapping of the scalar parameters
\begin{align}
	\overline{X(\epsilon,U,\alpha,\mu)} =
	X(\bar{\epsilon},\bar{U},\bar{\alpha},\bar{\mu})
	\label{eq:x_bar}
\end{align}
where $\bar{x}=-x$ for $x=\epsilon,U,\alpha,\mu$
while $T$ and $\Gamma$ are untouched and not explicitly indicated,
see~\cite{Ortmanns22a} for discussion of the parity operator $\P=(-1)^N$ and further details.
This suffices to introduce the appropriate duality-adapted variables
which simplify the \emph{analysis} of the solution --not just its computation--as much as possible.
The solution of Eqs.~\eqref{eq:master_equation}-\eqref{eq:I_E_energy_basis} reads
\begin{widetext}
		\begin{align}
			\Sket{\rho(t)} & =
			\big\{
			\tfrac{1}{4} \Sket{\one} + \brkt{A}_z \tfrac{1}{2} \Sket{A} + \brkt{p}_z \tfrac{1}{2}\Sket{p}
			\big\}
			\label{eq:state}\\
			&
			+
			\tfrac{1}{2} \Big[ \Sket{A} - \brkt{A}_{\bar{z}} \Sket{p}  \Big]
			e^{-\gamma_c t}
			\big[ \brkt{A}_{\rho_0} -\brkt{A}_{z} \big]
			+
			\Sket{p}
			e^{-\gamma_p t} \Big\{ \tfrac{1}{4} \big[ \brkt{p}_{\rho_0} -\brkt{p}_{z} \big]
			+ \tfrac{1}{2} \brkt{A}_{\bar{z}}  \big[ \brkt{A}_{\rho_0} -\brkt{A}_{z} \big]\Big\}
			\notag
			\\
			I_N(t)  &= \big( \gamma'_c +  \gamma'_s \brkt{A}_z \big)
			+ \gamma'_s e^{-\gamma_c t} \big[ \brkt{A}_{\rho_0} -\brkt{A}_{z} \big]
			\label{eq:charge_current}
			\\
			I_Q(t) & = 
			-\mu \big\{ \gamma'_c +  \gamma'_s \brkt{A}_z \big\}
			\notag\\
			&
			+ \Big\{ \tfrac{1}{2} \big( \delta_\A - U \brkt{A}_{\bar{z}} \big) \gamma_c -\mu \gamma'_s \Big\} e^{-\gamma_c t}
			\big[ \brkt{A}_{\rho_0} -\brkt{A}_{z} \big]
			+ U \gamma_p e^{-\gamma_p t} \Big\{ \tfrac{1}{4} \big[ \brkt{p}_{\rho_0} -\brkt{p}_{z} \big]
			+ \tfrac{1}{2} \brkt{A}_{\bar{z}}  \big[ \brkt{A}_{\rho_0} -\brkt{A}_{z} \big]\Big\}
			\label{eq:heat_current}
		\end{align}%
\end{widetext}%
where the energy current $I_E(t)$ is obtained by setting $\mu=0$ in $I_Q(t)$.
These expressions have the crucial advantage that they express the \emph{transient} dynamics in terms of \emph{stationary} expectation values of physical observables designated by duality.
These observables are
the superconducting \emph{polarization} $A=\sum_{\tau} \tau \ket{\tau}\bra{\tau}$
and the fermion \emph{parity} $p = (-\one)^{N}$
complemented by the trivial observable $\one$.
It is well-known that a quantum state can be expressed in such a set of observables with their time-dependent expectation values as coefficients,
\begin{align}
	\Sket{\rho(t)} & =
	\tfrac{1}{4} \Sket{\one} + \brkt{A}_{\rho(t)} \tfrac{1}{2} \Sket{A} + \brkt{p}_{\rho(t)} \tfrac{1}{2}\Sket{p}
	,
	\label{eq:bloch}
\end{align}
writing the state and the observables as supervectors:
\begin{subequations}%
	\begin{align}%
		\Sket{\one} & = \sum_{\tau}      \Sket{\tau} + 2 \Sket{1}
		\\
		\Sket{p}    & = \sum_{\tau}      \Sket{\tau} - 2 \Sket{1}
		\\
		\Sket{A}    & = \sum_{\tau} \tau \Sket{\tau}\ .
		\label{eq:A}
	\end{align}%
	\label{eq:observables}%
\end{subequations}% 
Here, however, the \emph{dynamics} can
be expressed in their expectation values in the actual \emph{stationary} state $z$,
$\brkt{\bullet}_z=\Sbraket{\bullet}{z}$
and, additionally, their values in the \emph{stationary} state $\bar{z}$ of a \emph{dual system},
$\brkt{\bullet}=\Sbraket{\bullet}{\bar{z}}$,
with inverted parameters~\eqref{eq:x_bar}
and the decay rates $\gamma_c$ and $\gamma_p$~\footnote
	{As explained in Ref.~\cite{Ortmanns22a}, using detailed balance
	it is possible to express the \emph{dual} stationary values $\brkt{A}_{\bar{z}}$ and $\brkt{p}_{\bar{z}}$
	in terms of the stationary values $\brkt{A}_z$,$\brkt{p}_z$.
	This interesting relation is, however, nonlinear,
	whereas the behavior of the dual stationary values is quite simple to physically understand
	since it corresponds to inversion of the interaction energy $\bar{U}=-U$, as explained Sec.~\ref{sec:invariants}.
}.
The dual quantities do not appear in the charge current formula~\eqref{eq:charge_current}
but \emph{only} enter through $\brkt{A}_{\bar{z}}$
in the transient part of the heat current~\eqref{eq:heat_current} via the energy current, a two-particle observable,
and contribute \emph{only} when the dot is interacting $U \neq 0$.
This highlights the relevance of fermionic duality to time-dependent many-particle transport.
We found that these expectation values can be expressed compactly
in terms of the two decay rates $\gamma_c$ and $\gamma_p$
and a signed transition-rate asymmetry $\gamma_s$~\cite{Ortmanns22a}:
\begin{subequations}
	\begin{alignat}{3}
		\brkt{A}_z
		& = -\frac{\gamma_s}{\gamma_c}
		&,\quad
		\brkt{p}_z
		& = 1- 2\frac{\gamma_c^2 - \gamma_s^2}{\gamma_p\gamma_c}
		\label{eq:Ap_expectation}\\
		\brkt{A}_{\bar{z}}
		& = + \frac{\bar{\gamma}_s}{\bar{\gamma}_c}
		,
		&
		\brkt{p}_{\bar{z}}
		& = 1- 2\frac{\bar{\gamma}^2_c - \bar{\gamma}^2_s}{\gamma_p \bar{\gamma}_c}
		\label{eq:Ap_expectation_dual}
	\end{alignat}%
	\label{eq:Ap_expectations}%
\end{subequations}%
defining $\bar{\gamma_s}= \gamma_s$ and $\bar{\gamma_c}=\gamma_p-\gamma_c$.
To compute the transport currents we need
besides $\delta$, $U$ and $\mu$
only two further coefficients $\gamma'_c$ and $\gamma'_s$.
The four variables $\gamma_c$, $\gamma_s$, $\gamma'_c$ and $\gamma'_s$ transform in a simple way~\cite{Ortmanns22a} under the duality mapping~\eqref{eq:x_bar} --explaining the simplicity of the solution formulas Eqs.~(\ref{eq:state})-(\ref{eq:heat_current})--
and are special linear combinations of the transport rates~\cite{Ortmanns22a}.
Pairwise the \emph{duality invariants} $\gamma_c$,$\gamma'_s$
and $\gamma_s$,$\gamma'_c$ behave similarly,
\begin{subequations}
	\begin{align}
		\gamma_c  & = \kappa_c  + \frac{\delta}{\delta_\A} \kappa'_s
		,&\quad
		\gamma'_c & = \kappa'_c + \frac{\delta}{\delta_\A} \kappa_s
		,\\
		\gamma_s  & = \kappa_s  + \frac{\delta}{\delta_\A} \kappa'_c
		,&	
		\gamma'_s & = \kappa'_s + \frac{\delta}{\delta_\A} \kappa_c
		,
	\end{align}%
	\label{eq:gammas}%
\end{subequations}%
%%%%%%%%%%%%%%%%%%%%%%%%%%%%%%%%%%%%%%%%%%%%%%%%%%%%%%%%%
\begin{figure}[t]
	\includegraphics[width=1.0\linewidth]{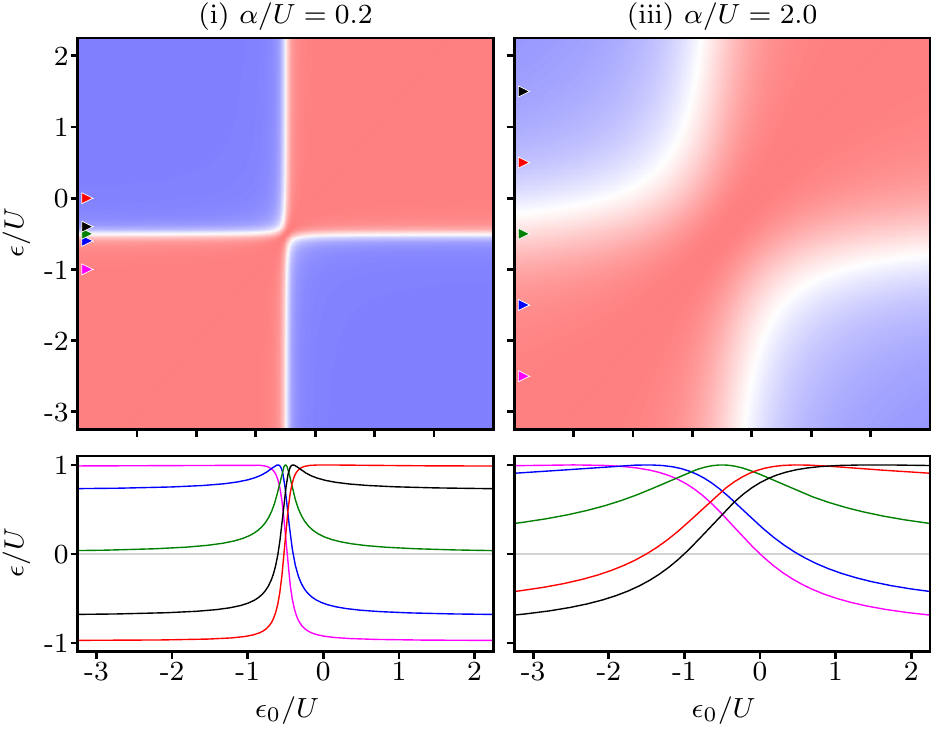}
	\caption{Overlap factor~\eqref{eq:theta} for the fast-switch initial-condition~\eqref{eq:Ap_initialize}
		depending on initial \emph{and} final gate voltage (red~$= +1$, blue~$=-1$).
		The horizontal line cuts are taken at positions indicated by arrows with corresponding color. We show the cases of weak (i) and strong (iii) pairing labeled as in Fig.~\ref{fig:thresholds} and following. 
		\label{fig:theta}
	}
\end{figure}%
%%%%%%%%%%%%%%%%%%%%%%%%%%%%%%%%%%%%%%%%%%%%%%%%%%%%%%%%%
since their ``components'' are the same elementary combinations of reservoir occupations (anti-)symmetrized over $\eta$ (particle type) and $\tau$ (energy state):
\begin{subequations}%
	\begin{align}%
		\kappa_c  & = \tfrac{1}{2}\gamma_p \sum_{\eta\tau}           f^{-\eta} (E_{\eta,\tau}-\mu)
		\\%, &\, 
		\kappa'_c & = \tfrac{1}{2}\gamma_p  \sum_{\eta\tau} \eta      f^{-\eta} (E_{\eta,\tau}-\mu)
		\\%\nonumber \\
		\kappa_s  & = \tfrac{1}{2}\gamma_p  \sum_{\eta\tau} \tau      f^{-\eta} (E_{\eta,\tau}-\mu)	
		\\%, & 
		\kappa'_s & = \tfrac{1}{2}\gamma_p  \sum_{\eta\tau} \tau \eta f^{-\eta} (E_{\eta,\tau}-\mu)\ .
	\end{align}%
	\label{eq:components}%
\end{subequations}%
These components separate the direct resonant \emph{broadening} effect of the superconductor [$\delta/\delta_\A$, see Eq.~\eqref{eq:Gamma_pm}]
from its indirect \emph{splitting} effect [via the Andreev levels $E_{\eta,\tau}$, see Eq.~\eqref{eq:energies_transition}]
allowing a clear analysis.

The state and transport evolution~\eqref{eq:state}-\eqref{eq:heat_current}
after the fast or slow switching initialization [Sect.~\ref{sec:initialization}]
is obtained by inserting for the \emph{initial} polarization and parity
\begin{align}
	\brkt{A}_{\rho_0} = \theta \brkt{A_0}_{z_0}
	,\quad
	\brkt{p}_{\rho_0} = \brkt{p}_{z_0}
	\label{eq:Ap_initialize}
\end{align}
Here $\brkt{A_0}_{z_0}$ and $\brkt{p}_{z_0}$ denote the stationary expectation values at gate voltage $\delta_0$ (instead of $\delta$) noting that we also have $A_0$ (instead of $A$): the polarization operator is itself gate-dependent like the energy eigenstates~[Eq.~\eqref{eq:A}].
As shown in App.~\ref{app:initial_condition}
for the slow switch we need to set $\theta=1$
whereas for the fast switch we have $\theta=$~Eq.~\eqref{eq:theta},
the overlap function plotted as function of the gate voltages in Fig.~\ref{fig:theta}.
This shows that
$\brkt{A}_{\rho_0}$ for the fast switch is obtained from the slow switch initial polarization $\brkt{A_0}_{z_0}$
by inverting its sign
(if $\theta/|\theta|=-1$, i.e., $\epsilon_0$ and $\epsilon$ lie on opposite sides of $-U/2$)
and contracting its magnitude
($|\theta| < 1$, i.e., $\epsilon_0$ or $\epsilon$ close to $-U/2$ on the scale $\alpha$).
The 
%%%%%%%%%%%%%%%%%%%%%%%%%%%%%%%%%%%%%%%%%%%%%%%%%%%%%%%%%
\begin{figure*}[t]
	\includegraphics[width=0.85\linewidth]{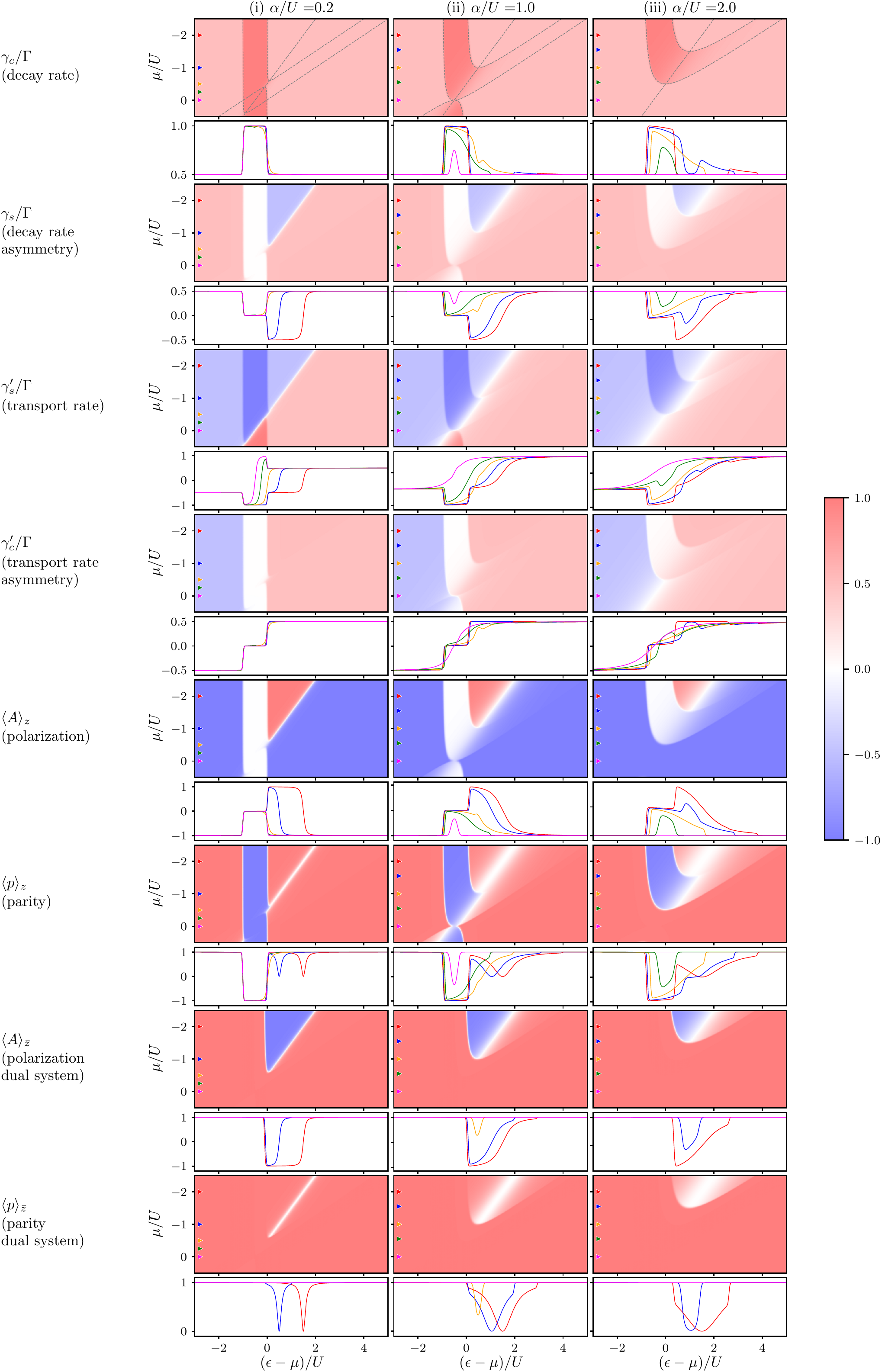}
	\caption{
		Invariants and stationary observables versus gate voltage $\epsilon-\mu$ and bias $\mu < 0$ 
		for $\alpha$ values of Fig.~\ref{fig:thresholds}.
		Line cuts in even rows are taken at arrows indicated in odd rows with same color.
	 	In row 1 the superconductor resonance (dashed line $\epsilon -\mu = -\mu - U/2$)
		and Andreev resonances (dashed curves where $\mu = E_{\eta,\tau}$ for $\eta, \tau=\pm$)
		%$\epsilon - \mu = \tfrac{1}{2} \big[ \tau \sqrt{ (2\mu-\eta U)^2 -\alpha^2} - U\big]-\mu$
		\label{fig:invariants}
	}
\end{figure*}
\cleardoublepage
\noindent
%%%%%%%%%%%%%%%%%%%%%%%%%%%%%%%%%%%%%%%%%%%%%%%%%%%%%%%%%
reason for this is that during a slow switch the quantum dot has time to exchange Cooper-pairs with the superconductor,
and thereby maintaining the polarization,
while during a fast switch this is not possible and can lead to a switch in polarization accounted for by the $\theta$ factor. 
Therefore, due to the superconductor, we have to carefully distinguish between
the \emph{initial} value $\brkt{A}_{\rho_0}$ of the dynamics
and the value \emph{before the switching} procedure $\brkt{A_0}_{z_0}$. Instead, the parity can never be changed by to the coherent Cooper-pair exchange with the superconductor.
%%%%%%%%%%%%%%%%%%%%%%%%%%%%%%%%%%%%%%%%%%%%%%%%%%%%%%%%%%%%%%%
\subsection{Duality invariant variables
	\label{sec:invariants}}
%%%%%%%%%%%%%%%%%%%%%%%%%%%%%%%%%%%%%%%%%%%%%%%%%%%%%%%%%%%%%%%

The analysis of the transport dynamics thus boils down to systematically understanding just four duality invariants~\eqref{eq:gammas}.
To facilitate later analysis we first establish their dependence on the physical parameters. 
They are plotted in the first four rows of Fig.~\ref{fig:invariants} 
as function of the experimentally controllable gate and bias voltage using the same color scale.
Since the metal probes the modification of the quantum dot's decay by the superconductor,
we plot all quantities versus $\epsilon-\mu$ and versus $\mu \leq 0$.

We first comment on the overall structure which will translate to later figures showing the stationary observables and transient current amplitudes.
Columns (i)-(iii)  show the development from weak to strong pairing relative to the interaction.
The Coulomb blockade (CB) regime ($-U < \epsilon -\mu < 0$) stands out in all quantities
as the vertical strip for $\alpha \ll U$
and is visible for any $\alpha$ at sufficiently large bias.
The superconductor resonance ($\epsilon = -U/2$) appears in these plots as the diagonal line
$ - \mu = \epsilon - \mu + U/2$ (dashed line in row 1)
hitting the CB regime at $\epsilon=0$ and bias $|\mu| = U/2$.
It broadens with increasing $\alpha$
since it derives from the effective couplings $\Gamma_\tau$, see Fig.~\ref{fig:thresholds}.

By contrast, the sharp changes occurring at gate voltages where
the conditions for resonances involving Andreev states are met, $E_{\eta,\tau} = \mu$ (dashed curves in row 1)
map out in a skewed fashion the spectrum in Fig.~\ref{fig:thresholds}.
Here and in the following we choose a sufficiently low temperature $T/U=0.015$
such that the Andreev features can be identified by sharp changes in color plots
and by clear steps in line cuts.
Their vertical asymptotes ($\epsilon - \mu = 0$ and $\epsilon-\mu+ U = 0$)
are the CB-like Andreev transitions.
The diagonal asymptotes have half the bias slope
[$(\epsilon - \mu)/2 =  - \mu$ and $(\epsilon-\mu + U )/2 =  - \mu$]
of the superconducting resonance.
These non-CB-like Andreev transitions become prominent
for strong pairing $\alpha \gtrsim U$
where the electron / hole components in the effective tunnel rates $\Gamma_\pm$
in the transition rates
start to have comparable weight, see Fig.~\ref{fig:thresholds}.
Importantly, this does not imply that Andreev transitions show up in all invariants and observables that depend on them:
this depends strongly on the quantity considered and on the bias regime.
The same applies to the superconductor resonance.

Specifically,
the rate of decay $\gamma_c=\sum_{\tau}W_{1,\tau}/2$ plotted in Fig.~\ref{fig:invariants}
equals an average of transition rates of the master equation~\eqref{eq:master_equation}.
Whenever some transition $\tau \to 1$ is enabled it will show up cumulatively in $\gamma_c$.
Since for any parameters ($U \geq 0$) $\gamma_c$ is bounded  as follows~\cite{Ortmanns22a}
\begin{align}
	\tfrac{1}{2}\gamma_p \leq \gamma_c \leq \gamma_p
	,
	\label{eq:gamma_c_bounds}
\end{align}
it always has a  magnitude that is comparable to the other, constant decay rate $\gamma_p=\Gamma$.
Transient measurements will thus always probe the interesting interplay of their two time scales.
We see in Fig.~\ref{fig:invariants}(i)
that for weak pairing  $\gamma_c$ can achieve the upper bound, $\gamma_c \approx \gamma_p$ (dark red) for a fixed gate-voltage interval,
whatever the bias.
This is the CB regime, where transitions to even-parity Andreev states are suppressed, while transitions from these
states to the odd parity state $|1)$ are enabled.
Here $\gamma_c$ is enhanced and the stationary system is singly occupied, $\Sket{z} \approx \Sket{1}$.
For strong pairing in Fig.~\ref{fig:invariants}(iii) 
$\gamma_c \approx \gamma_p/2$ (light red) is instead suppressed at low bias to its lower bound for any gate voltage
due to the pairing gap $\sim \alpha$ induced on the quantum dot, favoring the even-parity state $|-)$.
At intermediate bias, $\gamma_c$ interpolates smoothly between the two bounds
when varying $\epsilon-\mu$, interrupted by sharp steps whenever a transition involving an Andreev state is disabled or enabled.

The invariant $\gamma_s=\sum_{\tau} \tau W_{1,\tau}/2$ is the transition-rate asymmetry
taking on both negative and positive values
in Fig.~\ref{fig:invariants} to favor transitions starting from $\tau = +$ over transitions from $\tau = -$
or vice versa.
It is thus associated with the polarization of the even-parity energy states
[which is proportional to it, see Eq.~\eqref{eq:Ap_expectation}].
It is suppressed to zero (white) whenever $\gamma_c \approx \gamma_p$ in the first row of plots due to the bound~\cite{Ortmanns22a}
\begin{align}
	|\gamma_s| \leq \gamma_c,\gamma_p-\gamma_c
	\label{eq:gamma_s_bound}
\end{align}
It is also suppressed along the superconducting resonance
where $\Sket{z} \approx \tfrac{1}{2}\Sket{-}+\tfrac{1}{2}\Sket{+}$ is unpolarized (white)
interpolating between $\Sket{+}$ (blue) and $\Sket{-}$ (red) on either side.

The transport invariant $\gamma'_s=\sum_{\tau} \eta \tau W_{1,\tau}/2$ connects \emph{decay} of the polarization
to transient charge \emph{current}~\eqref{eq:charge_current}, see also Eq.~\eqref{eq:a} below.
In contrast to the decay rates, its dependence on $\epsilon-\mu$ in Fig.~\ref{fig:invariants} is roughly antisymmetric.
Unlike all other invariants, the superconductor resonance survives in the CB regime
where $\gamma'_s$ is not suppressed.

Finally, the invariant $\gamma'_c=\sum_{\tau} \eta W_{1,\tau}/2$ enters into the charge current,
which is also the only way in which it enters into the heat current~\eqref{eq:heat_current} via the contribution $-\mu I_N(t)$:
It is \emph{not} related to the energy current~\cite{Ortmanns22a}.
Fig.~\ref{fig:invariants} shows that it is likewise roughly antisymmetric:
At sufficiently high bias $|\mu| \gg \alpha$ it always inverts sign in the CB regime throughout which it vanishes, unlike $\gamma'_s$. For strong pairing it instead inverts sign at the superconducting resonance at low bias.

%%%%%%%%%%%%%%%%%%%%%%%%%%%%%%%%%%%%%%%%%%%%%%%%%%%%%%%%%%%%%%%
\subsection{Stationary local observables.
	\label{sec:stationary}}
%%%%%%%%%%%%%%%%%%%%%%%%%%%%%%%%%%%%%%%%%%%%%%%%%%%%%%%%%%%%%%%

Using the invariants is now straightforward to understand the parameter dependence
of expectation values of the observables $p$ and $A$ appearing in the solutions~\eqref{eq:state}-\eqref{eq:heat_current}.
They are plotted in the remaining rows of Fig.~\refeq{fig:invariants}
and can be explained using the first four rows.

The stationary polarization $\brkt{A}_z = -{\gamma_s}/{\gamma_c}$
is non-zero whenever the transition rate-asymmetry $\gamma_s$ is nonzero,
and is amplified in magnitude whenever the decay rate $\gamma_c$ is reduced.
One sees that for strong pairing
the system always prefers to be in the $\Sket{-}$ state, $\brkt{A}_z=-1$ (blue)
except for at high bias, where even the excited state $\Sket{+}$ can be favored, $\brkt{A}_z=1$ (red).
Likewise the stationary parity
$\brkt{p}_z  = 1- 2 (\gamma_c^2 - \gamma_s^2)/(\gamma_p\gamma_c)$
may be understood:
Having maximal positive parity, $\brkt{p}_z  = 1$ (red),
requires saturation of the bound~\eqref{eq:gamma_s_bound},
$|\gamma_s|=\gamma_c $,
which happens in the light red and light blue areas of the $\gamma_s$ plot.
Negative parity $\brkt{p}_z=-1$ (blue) instead requires $\gamma_s=0$ (white)
together with $\gamma_c=\gamma_p$ (dark red) and occurs only in the CB regime.
The blue areas clearly indicate the Coulomb blockade favoring single-occupation
characterized by odd parity $\brkt{p}_z=-1$ and $\brkt{A}_z=0$.
Zero parity $\brkt{p}_z = 0$ (white) occurs at the thermally sharp crossover to the CB-like regime (blue),
but also at superconducting resonance where $\gamma_s = 0$ (white) and $\gamma_c = \gamma_p/2$ (light red).
At the superconducting resonance all states are equally probable, leading to $\brkt{p}_z=0$ and $\brkt{A}_z=0$.

\begin{figure*}[t]
	\includegraphics[width=1.0\linewidth]{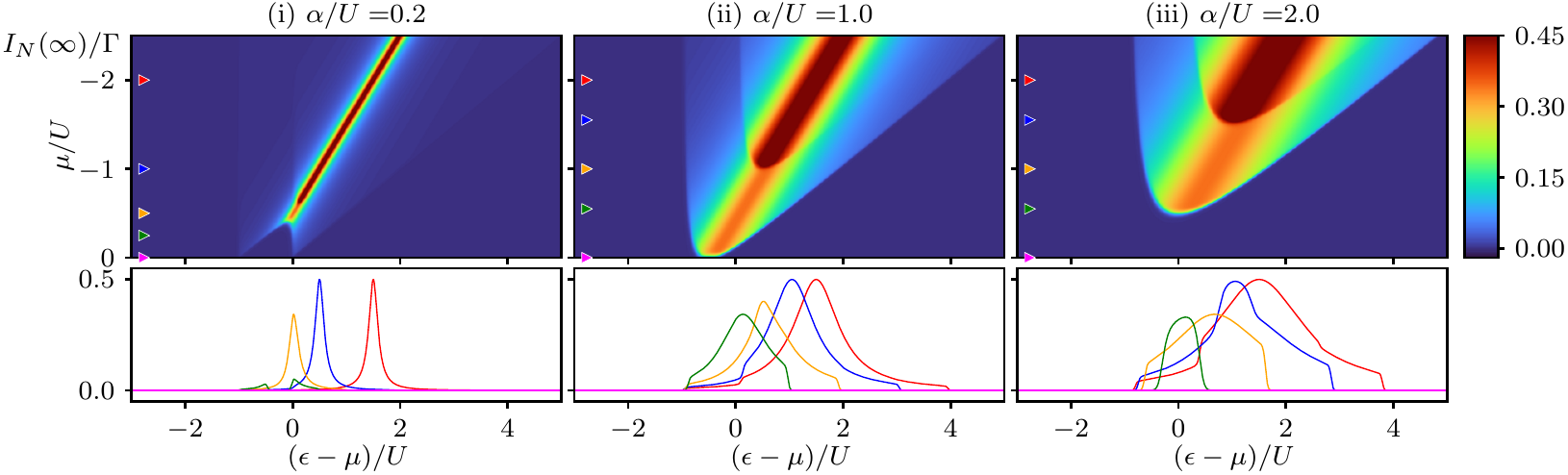}
	\caption{
		Stationary charge current $I_N(\infty)$ in units of $\Gamma$
		as function of gate and bias voltage.
		Parameters and conventions are the same as in Fig.~\ref{fig:invariants}.
		Note the faint Andreev resonances occurring at low bias $|\mu| \ll U$
		for $\alpha < U$.
	}
	\label{fig:stationary_current}
\end{figure*}
%%%%%%%%%%%%%%%%%%%%%%%%%%%%%%%%%%%%%%%%%%%%%%%%%%%%%%%

To analyze the solutions~\eqref{eq:state}-\eqref{eq:heat_current}
we furthermore need the expectation values of the same quantities
with respect to the dual stationary state $\bar{z}$
which are also plotted in Fig~\refeq{fig:invariants}.
These can be obtained in a similar way:
For the dual stationary polarization
$\brkt{A}_{\bar{z}}  = \bar{\gamma}_s/\bar{\gamma}_c$
we need to consider the plots of the dual invariants $\bar{\gamma_s}=\gamma_s$ and $\bar{\gamma_c}=\gamma_p-\gamma_c$.
The latter is simply obtained by recoloring dark red $\to$ white in the $\gamma_c$ plot.
Outside the CB regime if $\gamma_s \neq 0$,
this leads to the result that the dual polarization equals $\brkt{A}_{\bar{z}}=\pm 1$ (red / blue)
when $\gamma_s$ and $\gamma_c$ have the same (opposite) sign.
If $\gamma_s \to 0$ then $\brkt{A}_{\bar{z}}=0$
with one important exception that one should note carefully:
when $\gamma_c$ achieves its upper bound $\gamma_p$ (dark red)
in the CB regime
$\bar{\gamma_c}=\gamma_p-\gamma_c$ (not shown)
vanishes in \emph{exactly the same} way as $\gamma_s$.
As a result their ratio $\brkt{A}_{\bar{z}} \approx 1$ and shows \emph{no} signature of Coulomb blockade.

Likewise,
the dual stationary parity $\brkt{p}_{\bar{z}} = 1- 2 (\bar{\gamma}_c^2 - \bar{\gamma}_s^2)/(\gamma_p \bar{\gamma}_c)$
is now always positive and maximal, $\brkt{p}_{\bar{z}}= 1$
since for the absolute values the upper bound $|\bar{\gamma}_s|=|\gamma_s|=\bar{\gamma}_c=\gamma_p-\gamma_c$
is always achieved:
in these regions inverting the $\gamma_c$ plot colors (by swapping dark red with white) gives the $\gamma_s$ plot.
This holds true except at the superconducting resonance where $\brkt{p}_{\bar{z}}= -1$ 
since $\gamma_s$ vanishes but $\gamma_c$ does not.

The distinct behavior of these dual expectation values can be understood intuitively using fermionic duality
following Refs.~\cite{Schulenborg2016Feb,Schulenborg2020Jun}:
For the \emph{dual} system all energy parameters are inverted, in particular
the interaction $\bar{U}=-U<0$ is attractive, a well-known situation~\cite{Haldane1977Jan}.
As a result --even \emph{without} a superconductor-- the dual system favors even occupation,
i.e., it exhibits \emph{no} Coulomb-blockade stabilizing odd occupation
explaining why the stationary parity $\brkt{p}_{\bar{z}} = 1$ is even,
allowing $\brkt{p}_{\bar{z}} = 0$ only at the superconductor resonance.
Furthermore, the energy inversion of the dual model instead favors the (in the actual model) highest energy state, $|+)$, to be occupied,
explaining predominance of polarization $\brkt{A}_{\bar{z}}= 1$,
requiring a sufficiently large bias to access negative polarization $\brkt{A}_{\bar{z}}= -1$,
both opposite to the behavior of the actual system.

%%%%%%%%%%%%%%%%%%%%%%%%%%%%%%%%%%%%%%%%%%%%%%%%%%%%%%%%%
\subsection{Stationary charge and heat current}\label{sec:stationary_currents}
%%%%%%%%%%%%%%%%%%%%%%%%%%%%%%%%%%%%%%%%%%%%%%%%%%%%%%%%%

The stationary charge and heat current are proportional,
$
I_{Q}(\infty)  = - \mu I_{N}(\infty)
$,
since the stationary energy current vanishes,
$
I_E(\infty) =0
$.
This reflects that the Cooper pairs carrying the stationary particle current
do not transfer energy with respect to $\mu_\S=0$. It implies that the study of $I_Q(\infty)$ provides no advantage over $I_N(\infty)$
for probing the properties of the proximized dot~\cite{Ortmanns22a}, in contrast to the transient currents discussed later on. The parameter dependence of the stationary current can be understood in a similar way as the observables discussed in Sec.~\ref{sec:stationary}.
The stationary charge current
\begin{align}
	I_N(\infty)= \gamma'_c +  \gamma'_s \brkt{A}_z
	\label{eq:I_N_infty}
\end{align}
flowing into the metal probe for $\mu \leq 0$
is plotted in Fig.~\ref{fig:stationary_current},
see Refs.~\cite{Pala2007Aug,Governale2008Apr,Sothmann2010Dec,Droste2015Mar}, for similar results.
By Eq.~\eqref{eq:I_N_infty} its behavior is understood by combining the plots of $\gamma_c'$, $\gamma_s'$ and $\langle A\rangle_z$ in Fig.~\ref{fig:invariants}.

For strong pairing in Fig.~\ref{fig:stationary_current}(iii),
the current is suppressed below a bias voltage threshold for any gate voltage (magenta line cut):
here the nonzero transport invariants $\gamma'_c$ and $\gamma'_s$ are equal
but cancel out in Eq.~\eqref{eq:I_N_infty} since the stationary system is fully polarized, $\brkt{A}_z=-1$.
%~\footnote
%	{In terms of transitions, in this regime the state $\Sket{+}$ is with probability 1 in the stationary limit.
%	 Thus, no transitions take place and all transport halts.}
For weak pairing in Fig.~\ref{fig:stationary_current}(i),
the current is \emph{also} suppressed below a bias threshold but for a different reason:
even though transport invariant $\gamma'_s$ is now nonzero,
both $\gamma'_c$ and the polarization $\brkt{A}_z$ vanish 
in the CB regime.
As a result, the current displays a ``resonance with a hole'' (green, magenta line cuts).

By contrast, for strong pairing, the charge current in Fig.~\ref{fig:stationary_current} (iii)
shows a regular resonance peak at intermediate bias
as expected for the pair resonance (green, orange line cut)
but its tails are ``stepped'' due to a pair of Andreev transitions.
Fig.~\ref{fig:invariants} shows that at high bias $\gamma'_c$ is nonzero at the superconducting resonance
and cannot be canceled out in Eq.~\eqref{eq:I_N_infty}
by the product of $\brkt{A}_z$ and $\gamma'_s$ which are both zero at the resonance.
The broad resonance makes visible that another pair of Andreev transitions is activated, causing the curve to look like a ``stepped-pyramid''.

The line cuts also show that the gate-voltage dependence is clearly non-symmetric about the superconductor resonance ($\delta=0$)
and $I_N$ becomes symmetric in $\delta$ only for $\mu \to 0$.
By expressing the invariants in their components~\eqref{eq:components}
\begin{align}
	I_N(\infty)
	=
	\dfrac{\gamma'_c \gamma_c - \gamma'_s \gamma_s}{\gamma_c}
	= 
	\dfrac{\kappa_c \kappa'_c -\kappa_s \kappa_s'}
	     {\kappa_c +  \frac{\delta}{\delta_A} \kappa'_s}
	\, 
	\Big( 1- \frac{\delta^2}{\delta_A^2} \Big)
	\label{eq:Istat}%
\end{align}
we can in fact extract a symmetric Lorentzian-peak $\delta$-dependence $1- \delta^2/\delta_A^2=\alpha^2/(\delta^2+\alpha^2)$
of width $\alpha/2$
discussed in prior works~\cite{Pala2007Aug,Governale2008Apr,Sothmann2010Dec,Droste2015Mar}.
Our compact analytical expression brings out how this peak is nontrivially modified due to the bias asymmetry and stepped Andreev transitions.

%%%%%%%%%%%%%%%%%%%%%%%%%%%%%%%%%%%%%%%%%%%%%%%%%%%%%%%%%
\subsection{Transient charge current}\label{sec:charge_transient}
%%%%%%%%%%%%%%%%%%%%%%%%%%%%%%%%%%%%%%%%%%%%%%%%%%%%%%%%%

The transient charge current has single-exponential time-dependence
\begin{align}
	I_N(t)- I_{N}(\infty) & =  a e^{-\gamma_c t}
	\label{eq:charge_current_key}
	.
\end{align}
Due to this simple form
the charge-decay time scale is set by the invariant $\gamma_c$
and the initial value of the transient,
$I_N(0) -I_N(\infty)= a$,
completely characterizes the visibility of the transient charge current.
The amplitude,
\begin{align}
	a
	= \gamma'_s
	\big[ \brkt{A}_{\rho_0} -\brkt{A}_{z} \big]
	,
	\label{eq:a}%
\end{align}
is governed by transport invariant $\gamma'_s$
which has the same $\kappa$-components as the decay rate $\gamma_c$ [Eqs.~\eqref{eq:gammas}-\eqref{eq:components}].
Clearly, the transient charge current~\eqref{eq:charge_current_key}
\emph{only} probes the dot's initial \emph{excess polarization} $\brkt{A}_{\rho_0}$ relative to the final one, $\brkt{A}_{z}$.
The parameter dependence of $a$ for two possible switch scenarios is analyzed in  Sec.~\ref{sec:spectroscopy}-\ref{sec:discussion}.

%%%%%%%%%%%%%%%%%%%%%%%%%%%%%%%%%%%%%%%%%%%%%%%%%%%%%%%%%
\subsection{Transient heat current}\label{sec:heat_transient}
%%%%%%%%%%%%%%%%%%%%%%%%%%%%%%%%%%%%%%%%%%%%%%%%%%%%%%%%%

The transient heat current features a more intricate dependence on both the initial state and on time,
requiring some preliminary analysis before exploring its parameter dependence
in Sec.~\ref{sec:spectroscopy}-\ref{sec:discussion}.
This derives from the fact that the dot energy $H_\D$ is a two-particle quantity
--in contrast to charge--
allowing to probe the full correlated proximized dot state.

%%%%%%%%%%%%%%%%%%%%%%%%%%%%%%%%%%%%%%%%%%%%%%%%%%%%%%%%%%%%%%%
\subsubsection{Dependence on initial state}
%%%%%%%%%%%%%%%%%%%%%%%%%%%%%%%%%%%%%%%%%%%%%%%%%%%%%%%%%%%%%%%
To highlight the dependence on the initial state,
the transient heat current can be decomposed as
\begin{align}
	I_Q(t)-I_Q(\infty)
	=
	& I_{Q,A}(t) \big[ \brkt{A}_{\rho_0} -\brkt{A}_{z} \big]
	\notag
	\\
	& 
	+ I_{Q,p}(t) \big[ \brkt{p}_{\rho_0} -\brkt{p}_{z} \big]
	\label{eq:I_Q_initial}
\end{align}
into transient heat current contributions
flowing in response to an initial excess of quantities $A$ and $p$ on the proximized dot
relative to their final stationary values. The prefactors are given by
\begin{gather}
	%&
	I_{Q,A}(t)
	=
	\label{eq:IQA}
	\\
	%&
	\Big\{ \tfrac{1}{2} \Big( \delta_\A - 
	U \big( 1-\frac{\gamma_c}{\gamma_p} e^{-(\gamma_p - \gamma_c)t} \big)
	\brkt{A}_{\bar{z} }
	\Big)
	\gamma_c
	%\\
	%&
	 -\mu \gamma'_s
	\Big\} e^{-\gamma_c t}
	\notag
	,
	\\
	%&
	I_{Q,p}(t) =
	U \gamma_p e^{-\gamma_p t}
	.
\end{gather}
The transient heat current thus probes \emph{both} the decay of the initial excess of polarization \emph{and} parity.
It is possible to prepare a suitable energy-mixture $\Sket{\rho_0}$
for which these responses can be measured separately
using the basic physical initialization procedures~\eqref{eq:Ap_initialize} considered here.
Examples of such special slow or fast switches can be found by plotting the excess ratio $[\brkt{A}_{\rho_0} - \brkt{A}_{z} ]/[\brkt{p}_{\rho_0} - \brkt{p}_{z}]$ (not shown).

For initial states with $\brkt{A}_{\rho_0} = \brkt{A}_{z}$
the second term in Eq.~\eqref{eq:I_Q_initial} constitutes the full heat current with a response $I_{Q,p}(t)$
that depends \emph{only} on the interaction $U$ and the bare coupling $\gamma_p=\Gamma$.
It is independent of all other parameters,
in particular, the pairing $\alpha$ induced by the superconductor.
By contrast, for initial states with $\brkt{p}_{\rho_0} = \brkt{p}_{z}$ only the first term in Eq.~\eqref{eq:I_Q_initial} contributes which depends nontrivially on all parameters.
Interestingly, the sign of the interaction contribution to $I_{Q,A}(t)$ is reversed at a time
$t_1 \in [ \gamma_p^{-1}, \gamma_{c}^{-1}]$ given by
\begin{align}
	t_1 =\frac{ \ln(\gamma_p/\gamma_c) }{\gamma_p - \gamma_c}
	\label{eq:t1}
	.
\end{align}

%%%%%%%%%%%%%%%%%%%%%%%%%%%%%%%%%%%%%%%%%%%%%%%%%%%%%%%%%%%%%%%
\subsubsection{Non-monotonic dependence on time}\label{sec:nonmonotonicity}
%%%%%%%%%%%%%%%%%%%%%%%%%%%%%%%%%%%%%%%%%%%%%%%%%%%%%%%%%%%%%%%
To focus on the dependence on time we instead write
\begin{align}
	I_Q(t)- I_Q(\infty) & = a_c e^{-\gamma_c t}+a_p e^{-\gamma_p t}
	\label{eq:I_Q}
	,
\end{align}
with constant amplitudes given by
\begin{align}
%	\hspace{-0.17cm}
	a_c
	& =
	\Big\{ \tfrac{1}{2} \big( \delta_\A - U \brkt{A}_{\bar{z}} \big) \gamma_c -\mu \gamma'_s \Big\}
	\big[ \brkt{A}_{\rho_0} -\brkt{A}_{z} \big]
	\label{eq:a_c}%
	,
	\\
	\hspace{-0.17cm}
	a_p
	& = \gamma_p  U
	\Big\{
	\tfrac{1}{4} \big[  \brkt{p}_{\rho_0} -\brkt{p}_{z} \big]
	+ \tfrac{1}{2} \brkt{A}_{\bar{z}}  \big[ \brkt{A}_{\rho_0} -\brkt{A}_{z} \big]
	\Big\}
	\label{eq:a_p}
	\\
    & = \gamma_p U \Sbraket{p\bar{z}}{\rho_0}
    \label{eq:a_p_overlap}
    .
\end{align}%
Although the proximized dot is expected to have two timescales
based just on the Hilbert space dimension and constraints~\footnote
 	{Two real parameters remain based on the Liouville-space dimension 3 and 1 constraint of trace preservation.}
one cannot tell that both are actually \emph{relevant} except by computing their amplitudes.
Eqs.~\eqref{eq:a_c}-\eqref{eq:a_p} show that the \emph{interaction} $U$ is required for the second time-scale $\gamma_p$ to appear besides $\gamma_c$ ($a_p = 0$ for $U=0$)
extending a similar result of Ref.~\cite{Schulenborg2016Feb}.

The double-exponential transient heat current~\eqref{eq:I_Q} can exhibit \emph{non-monotonic} decay
since the amplitudes $a_c$ and $a_p$ may have opposite sign.
In fact, there are two types of such behavior.
First, the transient $I_Q(t)-I_Q(\infty)$ will pass through zero at time
\begin{align}
	t_0 = \frac{ \ln (-a_p/a_c)}{\gamma_p - \gamma_c}
	\label{eq:t0}
\end{align}
whenever this expression is positive.
In this case the fast parity decay ($\gamma_p$) has a larger amplitude than the slow charge decay ($\gamma_c$)
with opposite sign,
$0 \leq -a_c/a_p \leq 1$.
It will therefore initially push the transient through zero,
causing the heat current to intersects its stationary asymptote once, $I_Q(t_0)=I_Q(\infty)$,
before decaying to it from the opposite side.
The presence of such a zero implies that there is a local extremum---either a maximum or a minimum---at later time.

%%%%%%%%%%%%%%%%%%%%%%%%%%%%%%%%%%%%%
\begin{figure}[tb]
	\includegraphics[width=0.9\linewidth]{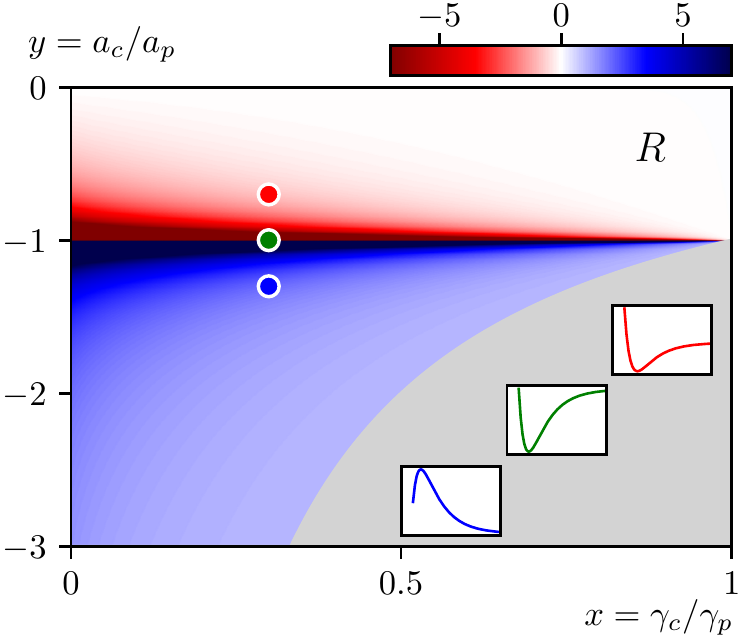}
	\caption{
		Possible values of nonmonotonicity quantifier $R$ [Eq.~\eqref{eq:extremum}]
		as function of the possible values of the ratio of decay rates and of amplitudes.
		The insets show the different non-monotonic profiles of the transient heat current
		as function of time
		for decay with a zero (red, $-1 \leq y < 0 $),
		for decay without a zero (blue $-1/x \leq y < -1$)
		and an intermediate case (green, $y= -1$).
		In the latter case the heat current is purely transient,
		(initial zero and finally zero).
	}
	\label{fig:nonmonotonicity}
\end{figure}
%%%%%%%%%%%%%%%%%%%%%%%%%%%%%%%%%%%%%

However,  a local extremum can also occur at a time which is positive,
\begin{align}
	t_2 =\frac{ \ln[-(\gamma_p a_p)/(\gamma_c a_c)] }{\gamma_p - \gamma_c}
	= t_0 + t_1
	\label{eq:t2}
	,
\end{align}
whenever $0 \leq (-\gamma_c a_c)/(\gamma_p a_p) \leq 1$.
In this case the initial \emph{rate of change} of the parity decay dominates that of the opposing charge decay,
$ (d/dt) a_p e^{-\gamma_p t}|_{t=0}
>-(d/dt) a_c e^{-\gamma_c t}|_{t=0}
$ without necessarily inducing the transient current to pass through zero.
Although $t_0$ is negative in case there is no passage of the transient through zero [Eq.~\eqref{eq:t0}]
it can still be compensated by $t_1$, the positive time~\eqref{eq:t1} at which the interaction contribution to $I_{Q,A}(t)$ reverses its sign [Eq.~\eqref{eq:IQA}], to achieve $t_2 \geq 0$.

We stress that the scales $t_0$, $t_1$ and $t_2$ characterizing the transient heat current profile in time are \emph{not}
any of the expected time scales of the state evolution ($\gamma_c^{-1}$, $\gamma_p^{-1}$)
but functions of these ($t_1$) and of their \emph{amplitudes} ($t_0$, $t_2$).
They emerge only due to nonzero interaction $U$.
A simple quantifier of a nontrivial interaction-induced
profile of the transient heat current
is provided by its extremal value at $t_2$ relative to its initial value:
\begin{subequations}%
\begin{align}%
	\frac{I_Q(t_2) -I_Q(\infty)}{I_Q(0) -I_Q(\infty)} =
	R\left( \frac{\gamma_c}{\gamma_p}, \frac{a_c}{a_p} \right)
	.
	\label{eq:R}
\end{align}%
which depends only on ratios of decay rates, $x=\gamma_c/\gamma_p$, and amplitudes, $y=a_c/a_p$, through the function ($y\leq 0$)
\begin{align}%
	R(x,y)=
	(-y x)^\frac{x}{1-x}  \frac{y}{y+1} (1-x)
	.
\end{align}%
\label{eq:extremum}%
\end{subequations}%
The possible values of $R$ are plotted in Fig.~\ref{fig:nonmonotonicity} as function of the two ratios
with possible line shapes.
The sign of $R$ indicates the type of non-monotonicity and its magnitude quantifies its degree~\footnote
	{For the special case of complete initial cancellation $a_c=-a_p$
	the value $|R|$ diverges, see Fig.~\ref{fig:nonmonotonicity}(c), inset with green curve.
	In this case one can use the extremal value of the transient relative to $a_p$
	to characterize the nonmonotonicity,
	$
	[ I_Q(t_2) -I_Q(\infty) ] / {a_p} =  x^{1/(1-x)} (1/x-1)
	$, depending only on $x=\gamma_c/\gamma_p$.}.
For $y \leq - 1/x$ (grey area) and $y \geq 0$ (not shown) there is no extremum.
We thus see that the initial value of transient heat current
$I_Q(0) -I_Q(\infty)=a_c + a_p$ does \emph{not} characterize the visibility
of the transient heat current because $a_c$ and $a_p$ can partially cancel.
Instead, one needs the extremal value relative to the initial value, given by $R$.

% LEAVE FOLLOWING EMPTY LINE HERE FOR mksubmit script !

%%%%%%%%%%%%%%%%%%%%%%%%%%%%%%%%%%%%%%%%%%%%%%%%%%%%%%%
\section{Transient spectroscopy}\label{sec:spectroscopy}
%%%%%%%%%%%%%%%%%%%%%%%%%%%%%%%%%%%%%%%%%%%%%%%%%%%%%%%

We now give a systematic overview of the predictions
as function of the experimentally controllable initial and final gate voltages,
while stepping through qualitatively distinct bias voltages (always choosing values $\mu \leq 0$ as before).
This analysis is done for the qualitatively different cases of weak ($\alpha = 0.2\, U$) and strong pairing ($\alpha = 2.0\, U$)
relative to the interaction, the parameters typically fixed in experiment.
We compare with the no-superconductor
%reference
case~\cite{Schulenborg2016Feb} which is first briefly reviewed.
Throughout this section we consider as before a fixed low temperature $T=0.015\,U$
to enable a clear analysis of the results:
We thus always have $\alpha \gg T$ such that the transitions involving Andreev states (\emph{split} by $\alpha$)
appear as sharp features
in contrast to the superconductor resonance (\emph{broadened} by $\alpha$).
In Sec.~\ref{sec:discussion} we provide a detailed explanation of all effects described here and we will discuss how they evolve with temperature $T$.

%%%%%%%%%%%%%%%%%%%%%%%%%%%%%%%%%%%%%%%%%%%%%%%%%%%%%%%
\subsection{No superconductor ``$\alpha \to 0$''}\label{sec:no_pairing}
%%%%%%%%%%%%%%%%%%%%%%%%%%%%%%%%%%%%%%%%%%%%%%%%%%%%%%%

\begin{figure}[t]
	\includegraphics[width=1.0\linewidth]{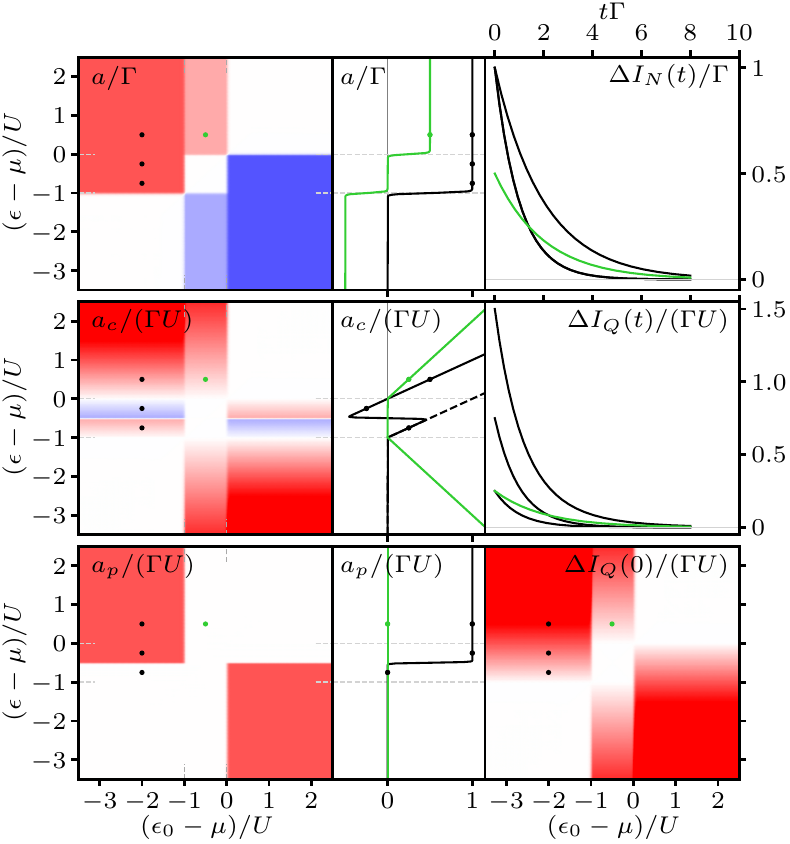}
	\caption{
		Quantum dot probed by metal (no superconductor).
		Left: Amplitudes for charge, $a$, and heat current, $a_c$ and $a_p$,
		as function of gate-voltage pair $(\epsilon_0,\epsilon)$ defining a switch.
		Center: Vertical line cuts through the points with the corresponding color.
		Right: Full transients at these points and 
		initial transient heat current $\Delta I_Q(0)=a_c+a_p$ versus gate voltages.
		Gray dashed lines indicate the gate voltages bounding the
		Coulomb blockade regime ($-U < \epsilon < 0$).
	}
	\label{fig:amplitudes_no_pairing}.
\end{figure}

We start from the well-understood no-superconductor case 
studied in Ref.~\cite{Schulenborg2016Feb}
which is included in our results~\cite{Ortmanns22a} by considering the \emph{fast} switch
and \emph{formally} taking $\alpha \to 0$ in Eqs.~\eqref{eq:charge_current}-\eqref{eq:heat_current}.
All possible transient charge- and heat-current experiments
can be analyzed by plotting
the amplitudes $a$, $a_c$ and $a_p$
for all initial and final gate-voltage pairs
$\epsilon_0-\mu$ and $\epsilon-\mu$
as in Fig.~\ref{fig:amplitudes_no_pairing}.
Experimentally, these amplitudes can be extracted by fitting the transient currents.
In a second step, we combine these with the dependence of the decay rates $\gamma_c$ and $\gamma_p$
on \emph{final} gate voltage $\epsilon-\mu$
to analyze the possible directly accessible transient currents
$\Delta I_N(t) = I_N(t)-I_N(\infty)$ and $\Delta I_Q(t)= I_Q(t)-I_Q(\infty)$.
These are plotted  in Fig.~\ref{fig:amplitudes_no_pairing} as function of time
for a selection
%of indicated
gate-voltage points.

In the no-superconductor case, we see in Fig.~\ref{fig:amplitudes_no_pairing}
that for switches from low to high gate voltage,
expelling electrons from the dot,
the charge amplitude $a =\Delta I_N(0)$ takes constant positive values (red)
on well-defined plateaus with sharp, thermally broadened boundaries.
For these switches the transient charge current ($=$particle current) monotonically decays into the metal
whereas for opposite switches it flows out (blue).

%%%%%%%%%%%%%%%%%%%%%%%%%%%%%%%%%%%%%%%%%%%%%%%%%%%%%%%
%%%%%%%%%%%%%%%%%%%%%%%%%%%%%%%%%%%%%%%%%%%%%%%%%%%%%%%

The heat amplitudes $a_c$, $a_p$ show linear and constant behavior, respectively,
as function of the final value $\epsilon-\mu$ (black line cut in Fig.~\ref{fig:amplitudes_no_pairing}),
except for a surprising thermally-broadened jump at $\epsilon-\mu = -U/2$.
(not to be confused with the superconductor resonance $\epsilon=-U/2$ later on).
When added together to obtain the initial transient $a_c + a_p = \Delta I_Q(0)$ (black dashed line cut)
these opposite jumps cancel out to give a smooth dependence on $\epsilon-\mu$
[lower right panel in Fig.~\ref{fig:amplitudes_no_pairing}].
At later times $\Delta I_Q(t)$ develops a kink in the $\epsilon-\mu$ dependence [not shown, see Sec.~\ref{sec:temperature}, Fig.~\ref{fig:temperature} (a)].
The transient heat current $\Delta I_Q(t)$ decays monotonically with $t$
for all switches, just like $\Delta I_N(t)$,
despite the occurrence of opposite signs of $a_c$ (blue) and $a_p$ (red) (see Sec.~\ref{sec:heat_transient}).
This is always the case for thermal broadening much smaller than the interaction ($T \ll U$).
Here, without superconductor, a nonmonotonic transient heat current is possible
but only when the thermal broadening is comparable with interaction ($T \sim U$)
which corresponds to cooling the metal noting that $I_Q(\infty)=0$.
This is discussed in App.~\ref{app:reversal} and in Ref.~\cite{Vanherck2017Mar} for two biased normal contacts.

Finally,
when comparing the results with and without superconductor
one should keep in mind that within the range of validity of our approach
even the case of weak pairing $\alpha \ll U$
is \emph{not} a ``weak perturbation'' continuously connected to the no-superconductor case:
Clearly, we cannot send $\alpha \to 0$ without violating the assumption $\Gamma \ll \alpha$ made in Sec.~\ref{sec:model}.
This is potentially confusing since one can extract the no-superconductor result
from our formulas by \emph{formally} sending $\alpha \to 0$ for the \emph{fast switch}
when carefully canceling \emph{discontinuous} contributions~\footnote
	{In \cite{Ortmanns22a} this is analytically verified
	using that in our duality-invariant formulation discontinuities are ``automatically'' collected.
	By contrast, for the slow switch, not well-defined for $\alpha \to 0$,
	the results do \emph{not} connect to result of Ref.~\cite{Schulenborg2016Feb}
	as expected (there is no anticrossing state to follow adiabatically).
}, see the detailed discussion in Ref.~\cite{Ortmanns22a}.

%%%%%%%%%%%%%%%%%%%%%%%%%%%%%%%%%%%%%%%%%%%%%%%%%%%%%%%
\subsection{Weak pairing ($\alpha < U$)}\label{sec:weak_pairing}
%%%%%%%%%%%%%%%%%%%%%%%%%%%%%%%%%%%%%%%%%%%%%%%%%%%%%%%

We now turn to the effects introduced by the superconductor
focusing first on the amplitudes of the transient response.
The pairing $\alpha > 0$ enters in two ways:
through the invariants (combinations of transition rates)
and the choice of fast vs. slow switch (the initial state).
We thus need to consider the amplitudes of the transients for both fast and slow switch
since they lead to different results.
%%%%%%%%%%%%%%%%%%%%%%%%%%%%%%%%%%%%%%%%%%%%%%%%%%%%%%%
\begin{figure*}[t]
	\includegraphics[width=1.0\linewidth]{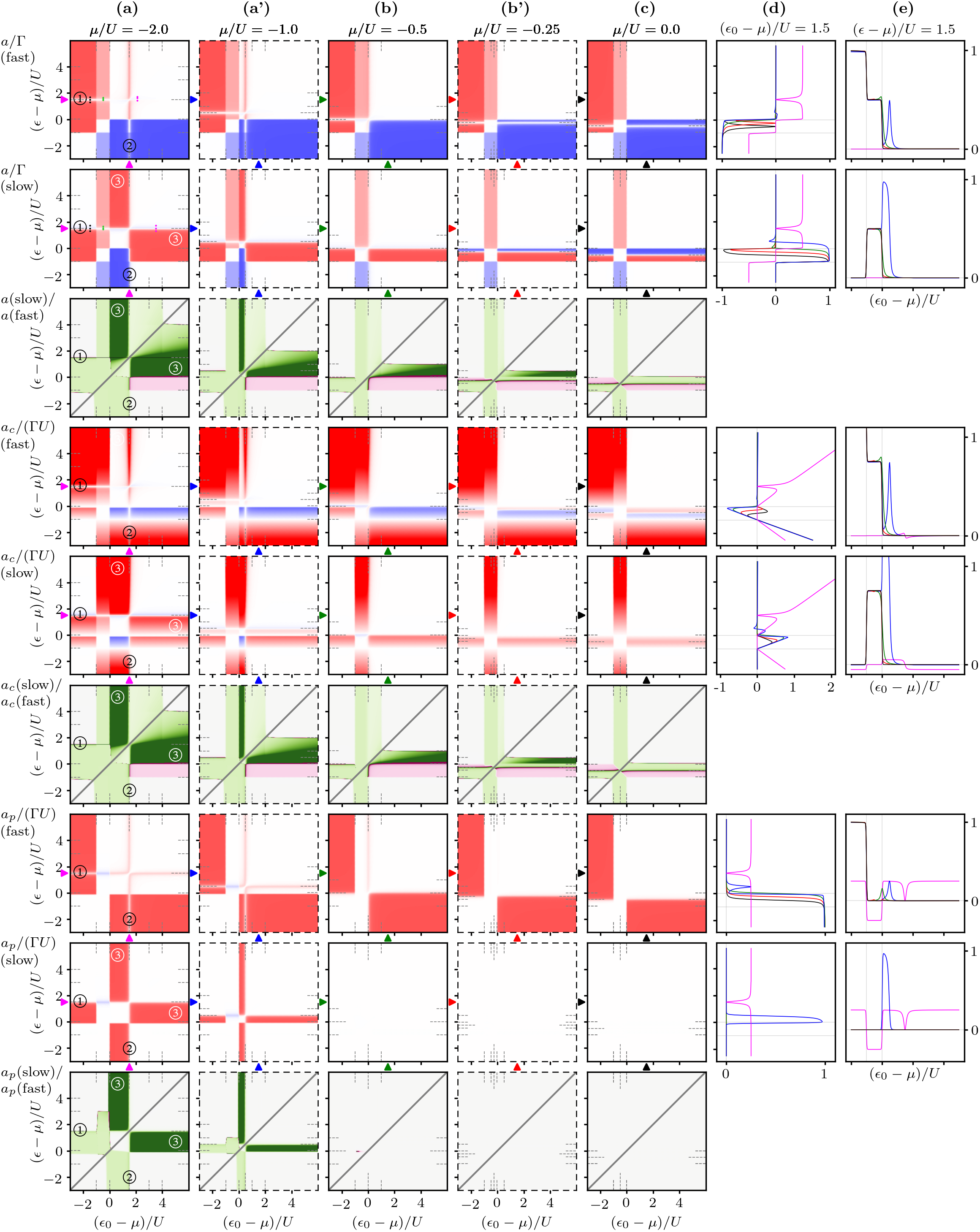}
	\caption{
		Weak pairing ($\alpha = 0.2\, U$):
		Charge (heat) amplitudes $a$ ($a_c$ and $a_p$) in units of coupling $\gamma_p=\Gamma$ ($\times$ interaction energy $U$),
		see main text for description of the layout.
		Vertical (horizontal) line cuts in column (d) [(e)]
		share the vertical (horizontal) gate voltage axis with columns (a)-(c).
		Horizontal (vertical) arrowheads in (a)-(c) indicate where vertical (horizontal) line cuts in (d) [(e)] are taken.
		The shorter (longer) dashed lines indicate the gate-voltage positions of the Andreev (superconductor) resonances.
		At the diagonal we trivially don't switch and the ratio of amplitudes (0/0) is not defined (grey line in rows 3,6,9).
	}
	\label{fig:amplitudes_weak_pairing}
\end{figure*}%
%%%%%%%%%%%%%%%%%%%%%%%%%%%%%%%%%%%%%%%%%%%%%%%%%%%%%%%
\begin{figure*}[t]
	\includegraphics[width=1.01\linewidth]{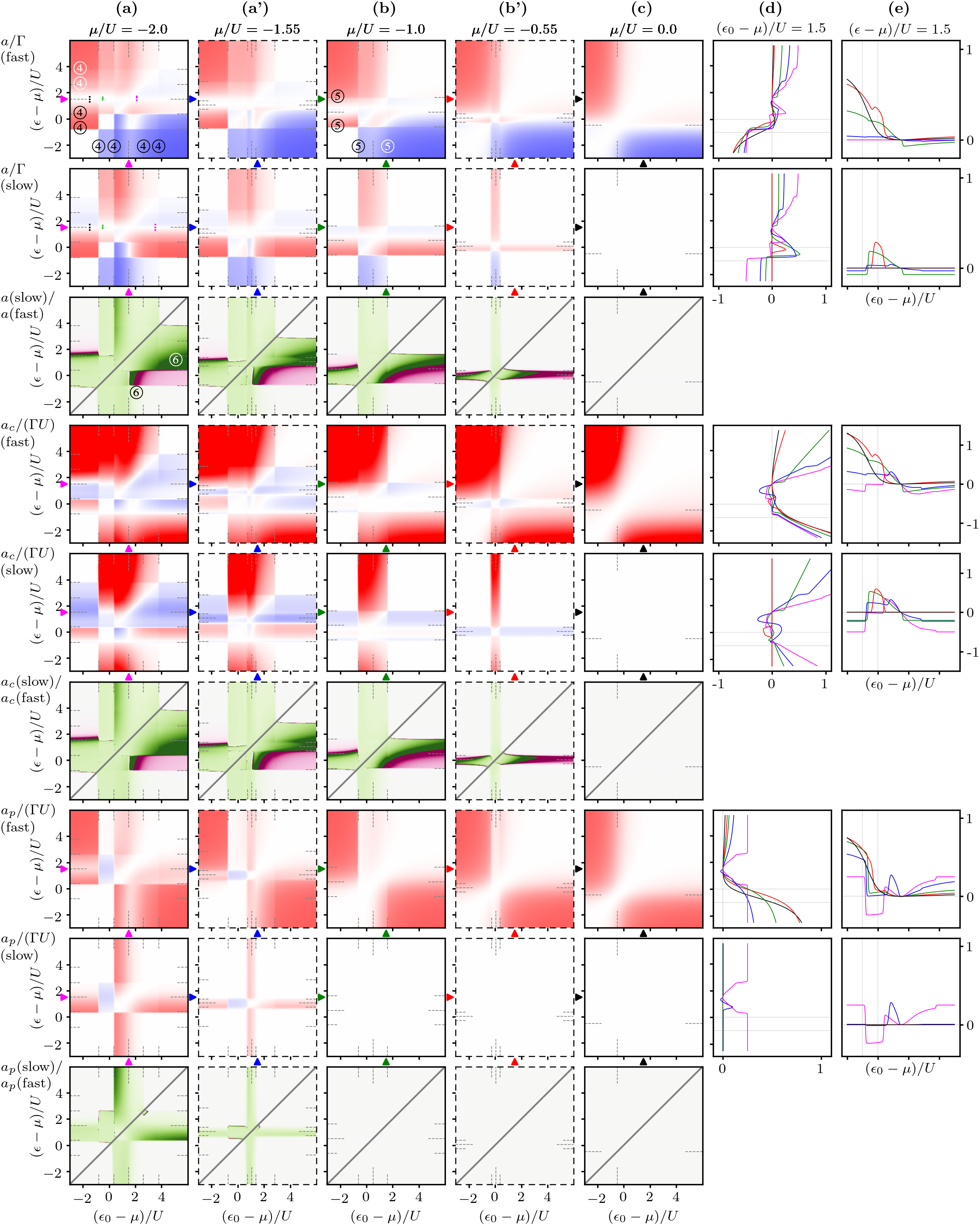}
	\caption{
		Strong pairing ($\alpha = 2.0\, U$): See caption of Fig.~\ref{fig:amplitudes_weak_pairing} and the main text.
	}
	\label{fig:amplitudes_strong_pairing}
\end{figure*}%
\cleardoublepage
\noindent
%%%%%%%%%%%%%%%%%%%%%%%%%%%%%%%%%%%%%%%%%%%%%%%%%%%%%%%
Together with their ratio, they are plotted in Fig.~\ref{fig:amplitudes_weak_pairing} in groups of three rows to facilitate comparison.
Furthermore,
the presence of the biased superconductor leads to stationary non-equilibrium currents:
for our choice $\mu \leq 0$ the stationary particle current always flows into the metal,
and heat current \emph{typically} also does, i.e.,
$I_N(\infty),I_Q(\infty) > 0$.
As for the no-superconductor case
the \emph{transient} heat current may reverse in time for $T \sim U$
but now also for $T \ll U$
(see Sec.~\ref{sec:heat_transient} and Sec.~\ref{sec:reversal}).
Thus, in Fig.~\ref{fig:amplitudes_weak_pairing},
a positive heat current amplitude (red) \emph{only} indicates that it \emph{initially favors}
a \emph{transient} heat current in the direction of the stationary flow;
vice versa, a negative amplitude (blue) indicates that an initial transient is favored which flows against the stationary heat current.
The bias $\mu$ also modifies how these transient currents approach their stationary values.
In columns (a)-(c) of Fig.~\ref{fig:amplitudes_weak_pairing}
we thus have to consider the qualitatively different biasing situations
that we identified earlier in Fig.~\ref{fig:thresholds}.
Additional cases (a$'$) and (b$'$) --lying closer to the thresholds of Fig.~\ref{fig:thresholds}-- are shown in Fig.~\ref{fig:amplitudes_weak_pairing} (dashed panel borders)
for better interpolation of the $\mu$-dependence.

%%%%%%%%%%%%%%%%%%%%%%%%%%%%%%%%%%%%%%%%%%%%%%%%%%%%%%%
\subsubsection{Fast switch}\label{sec:weak_pairing_fast_switch}
%%%%%%%%%%%%%%%%%%%%%%%%%%%%%%%%%%%%%%%%%%%%%%%%%%%%%%%
We first discuss results for high bias plotted in column~(a) of Fig.~\ref{fig:amplitudes_weak_pairing},
where the various contributions are clearly separated. For the weak pairing regime we are focusing on now, this corresponds to $|\mu| > U/2$ (see Fig.~\ref{fig:thresholds}).
The fast-switch amplitudes (row 1, 4, 7) show
the signature of the no-superconductor case (the pattern of Fig.~\ref{fig:amplitudes_no_pairing}).
This is modified by the superconductor by a pronounced cross-shaped resonance.

\emph{Final superconductor resonance (horizontal)}~\marker{1}:
For any fast switch where the \emph{final} gate voltage is resonant with the superconductor ($\epsilon=-U/2$)
there is a horizontal feature of small width $\alpha < U$ starting at marker~\marker{1}. 
For the charge amplitude $a$ this is a dip,
a suppression of $a$ relative to the no-superconductor background.
It goes all the way to zero implying zero transient charge current, $I_N(t)-I_N(\infty) = 0$ for $t\geq 0$.
Likewise, one heat current amplitude is completely suppressed at this horizontal feature, $a_c \approx 0$.

As a result,
if the final gate voltage is resonant with the superconductor
the transient heat current is entirely given by the remaining \emph{two-particle} contribution
$I_Q(t)-I_Q(\infty) \approx a_p e^{-\gamma_p t}$
originating from the energy-current part of~Eq.~\eqref{eq:heat_current}.
The direction of this remaining transient current varies horizontally along this feature
as function of the initial gate voltage:
For $\epsilon_0-\mu < -U$ the amplitude $a_p$ also shows a positive dip (light red)
relative to the no-superconductor background (dark red) but it is not fully suppressed.
For $\epsilon_0 -\mu > 0$, instead there is a positive peak where $a_p$ takes on the same value (light red)
on top of the zero background (white)
[except around $\epsilon=\epsilon_0$ where trivially $a_p=a_c=a=0$ since we don't switch].
This corresponds to a transient heat current \emph{with} the stationary flow into the metal. 
By contrast, in the initial CB regime $-U < \epsilon_0 -\mu < 0$,
$a_p$ shows a negative peak (blue) on the zero background (white)
corresponding to a transient heat current \emph{against} the stationary flow.
The values taken by the different amplitudes, $a$, $a_c$, and $a_p$ at this horizontal resonance in panels (a) are shown by a horizontal line cut (magenta) in panels (e) taken along this feature.

\emph{Initial superconductor resonance (vertical)}~\marker{2}:
For any fast switch where the \emph{initial} gate voltage is resonant with the superconductor ($\epsilon_0=-U/2$),
the no-superconductor background is also interrupted but in a different way.
This vertical feature starts at marker~\marker{2} in Fig.~\ref{fig:amplitudes_weak_pairing}(a).
The values of the amplitudes along this vertical feature are shown in panels (d) by a vertical line cut (magenta).
In this case, the charge amplitude $a$ shows a dip in the (now negative) no-superconductor background (blue)
suppressing it either to zero ($-U < \epsilon-\mu < 0$) or to only half its value ($\epsilon-\mu < -U$).
In the remaining case ($\epsilon - \mu > 0$)
there is now a positive (red) peak 
on the zero no-superconductor background (white)
[except trivially around $\epsilon=\epsilon_0$, the crossing with the vertical resonance].
The heat amplitude $a_c$ likewise shows only partial suppression
when the background is nonzero ($\epsilon-\mu < -U$)
and a positive resonant peak (red)
%of either sign (red/blue)
when the background is zero ($\epsilon - \mu > 0$).
\emph{Only} in the final CB regime ($-U<\epsilon-\mu<0$) do we have $a_c \approx 0$
and thus again a ``pure'' two-particle transient heat current.
This remaining contribution and its amplitude $a_p$ are
strictly positive
along the entire vertical resonance
[except trivially at $\epsilon=\epsilon_0=0$], see vertical line cut (magenta) in (d).
The apparent \emph{lack} of any effect of the interaction $U$ on the value of $a_p$ along the vertical initial resonance
--unlike the horizontal final resonance (blue CB feature)-- is surprising and will be explained in Sec.~\ref{sec:discussion}.

We now follow these resonances as we step through the qualitatively distinct bias values in columns (a)-(c) in Fig.~\ref{fig:amplitudes_weak_pairing}.
We see that the vertical initial superconductor resonance~\marker{2} in $a$ disappears from all amplitudes
when it hits the initial CB-regime ($-U < \epsilon_0-\mu <0$)
as one might have expected since interaction dominates pairing ($\alpha \ll U$).
One might also expect that the horizontal resonance disappears
but this does \emph{not} happen.
For the charge current amplitude $a$ the dip~\marker{1} persists,
$a=0$ at $\epsilon=-U/2$, no matter how small the nonzero pairing $\alpha$ is~\footnote
	{In the limit $\alpha \to 0$ the experimental measurement time for resolving this feature diverges, see Fig.~\ref{fig:switch} since our results are valid only for $\alpha\gg\Gamma$.
	 As mentioned, this is formally reflected by the results becoming non-continuous functions of gate voltages.
 	}.
It moves through the initial CB regime $-U < \epsilon_0-\mu <0$
as we vary the bias $|\mu|$ down to zero.

Similarly, the heat current amplitude $a_c$ shows a zero with sign change
which slides through this regime when lowering the bias.
Also for amplitude $a_p$ the superconductor resonance is not absent:
Upon lowering $|\mu|$, when the superconductor resonance hits the CB regime,
the plateau-step (red) at $\epsilon-\mu=0$ in (b), it starts dragging this step downward in (b$'$),
until the step reaches $\epsilon=-U/2$ for $\mu=0$ in (c).
This sliding of the step through the CB regime is the signature of the superconductor resonance in $a_p$.
Altogether, as emphasized at the end of Sec.~\ref{sec:no_pairing}, for low bias the weak pairing results, $\alpha \ll U$, are \emph{not} a slight perturbation of the no-superconductor result despite the first appearance of $a_c$ and $a_p$ at $\mu=0$.
The same conclusion follows at high bias $|\mu | > U$~\footnote
	{In Fig.~\ref{fig:amplitudes_weak_pairing}(a) amplitudes $a$ and $a_p$ resemble
	 the no-superconductor result in Fig.~\ref{fig:amplitudes_no_pairing} for gate voltages far away from the superconductor resonance. However, it is now the heat current amplitude $a_c$ that looks completely different.}.
The subtle reason for the close resemblance
 of $a_c$ and $a_p$ at $\mu=0$ in Fig.~\ref{fig:amplitudes_weak_pairing}(c)
to the no-superconductor result in Fig.~\ref{fig:amplitudes_no_pairing}
is that in the latter case $a_c$ and $a_p$ are \emph{already suppressed} at $\epsilon -\mu = -U/2$ as mentioned in Sec.~\ref{sec:no_pairing}.
This hides the signature of the superconductor resonance at $\epsilon = -U/2$ which for $\mu=0$ happens to occur at the same position.

%%%%%%%%%%%%%%%%%%%%%%%%%%%%%%%%%%%%%%%%%%%%%%%%%%%%%%%
\subsubsection{Slow switch}\label{sec:weak_pairing_slow_switch}
%%%%%%%%%%%%%%%%%%%%%%%%%%%%%%%%%%%%%%%%%%%%%%%%%%%%%%%
We now inspect the results in Fig.~\ref{fig:amplitudes_weak_pairing} for the slow switch (rows 2, 5, 8)
guided by the ratio of the slow and fast switch results plotted in the same figure (rows 3, 6, 9).
We see that in extended regions the amplitudes for the slow switch are \emph{suppressed} (white areas)
where those for the fast switch are not.
However, marker~\marker{3} indicates extended gate voltage regimes in which also the reverse can happen,
i.e., the slow switch amplitudes \emph{survive} while the fast switch amplitudes are suppressed (dark green / red in ratio plots).
For high bias (a)-(a$'$) we observe the following:
whenever for the fast switch the vertical superconductor resonance appears as a peak  ($\epsilon_0=-U/2$) rising from the zero background, the slow switch result continues as a plateau  ($\mu < \epsilon_0 < -\mu - U/2$ noting $-\mu \geq U/2$).
Instead, whenever for the fast switch the superconductor resonance appears  as a dip (vertically/horizontally),
dropping to zero from the non-zero background,
the slow switch result continues as a zero plateau ($\epsilon_0 > \mu $ / $\epsilon > \mu$).
Despite these differences,
precisely along both the vertical and horizontal resonance in panels (a)
[magenta line cuts in panels (d)-(e)]
the slow and fast switch results for $a$ and $a_p$ coincide.
For $a_c$ this holds only along the vertical resonance.

For $a$ and $a_c$ it may also happen that the slow and fast switch have comparable amplitudes but with the \emph{opposite sign} (light red in ratio plots).
At low bias at the horizontal superconductor resonance
in $a$ there is a sign change as function of $\epsilon$ for the slow switch, but not for the fast switch.
For $a_c$ this the other way around.
Note in particular that $a_c$ does not even have a zero at the resonance for the slow switch.
As a result
for $\mu=0$
the slow switch result for these amplitudes
does not even qualitatively resemble the no-superconductor result,
even for small $\alpha \ll U$, in contrast to the case of the fast switch (where, as discussed, ``only'' a superconductor resonance feature of small width remains).
Despite the large areas of identical response (light green in ratio plots) there are thus significant differences
between the two types of transient experiments,
as intuitively expected, due to the possibility of exchanging Cooper pairs with the superconducting contact during a slow switch as discussed in detail in Sec.~\ref{sec:discussion_weak_pairing}.

Finally, the completely white panels for $a_p$ highlight that there are bias regimes where the slow-switch response is  \emph{entirely thermally activated} even though the induced gap $\alpha$ is small ($\alpha < U$).
The amplitudes are exponentially suppressed and only become visible in these panels when temperature is increased,
resulting in non-monotonic $T$ dependence%
, see Sec.~\ref{sec:discussion}.
For the low temperature chosen here, $a_p$ is essentially zero for all possible switches
until both Andreev energy thresholds become accessible for bias $|\mu| > U$
going from (b) $\to$ (a$'$).
This differs strongly from the fast-switch result for $a_p$.

%%%%%%%%%%%%%%%%%%%%%%%%%%%%%%%%%%%%%%%%%%%%%%%%%%%%%%%
\subsection{Strong pairing ($\alpha > U$)}\label{sec:strong_pairing}
%%%%%%%%%%%%%%%%%%%%%%%%%%%%%%%%%%%%%%%%%%%%%%%%%%%%%%%

So far only the two CB-like Andreev transitions
had an effect on the amplitudes, both for the fast and slow 
switch.
These show up as steps in $a$ and steps or zeros in $a_c$.
The other two Andreev transitions (Fig.~\ref{fig:invariants}) 
had no noticeable effect
despite being sharply defined by the low temperature ($T \ll \alpha$).
This changes when the pairing $\alpha$ dominates over the interaction $U$.

%%%%%%%%%%%%%%%%%%%%%%%%%%%%%%%%%%%%%%%%%%%%%%%%%%%%%%%
\begin{figure*}[t]
	\includegraphics[width=1.0\linewidth]{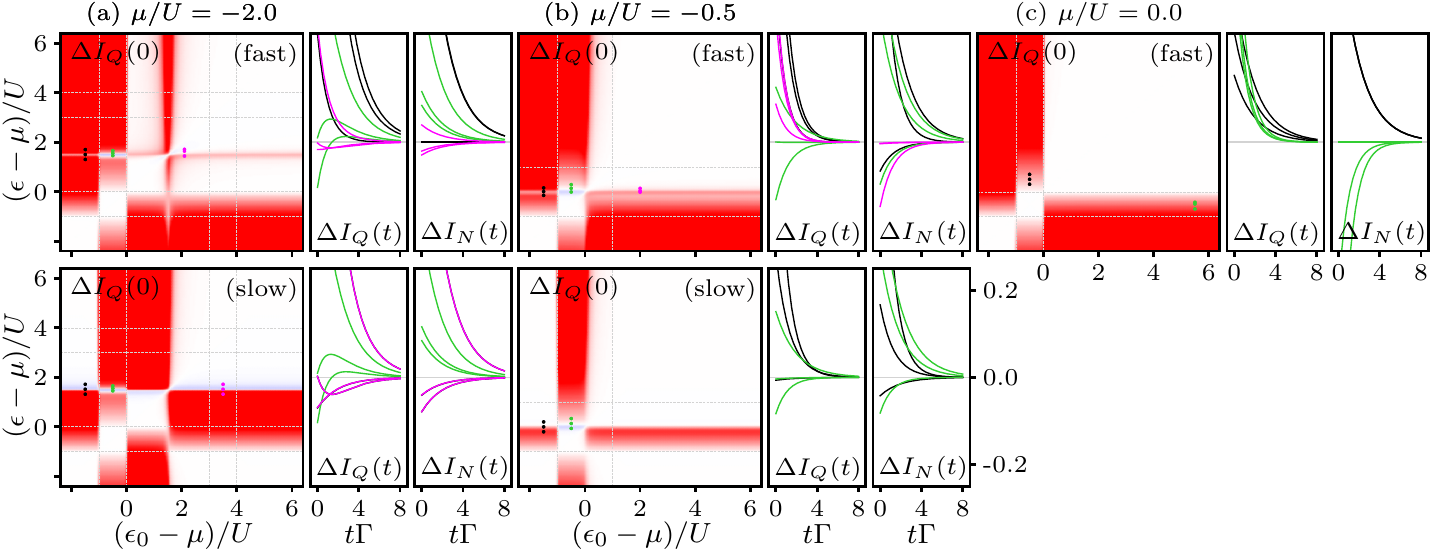}
	\caption{Weak pairing ($\alpha = 0.2\, U$):
		Transient heat current $\Delta I_Q(0)=I_Q(0)-I_Q(\infty)$ at $t=0$ in units $\Gamma \, U$
		as function of gate voltages for fast switch (row 1) and slow switch (row 2)
		where for the slow switch case (c) is left out (transients entirely thermal activated, essentially zero at this value of $T$, see Sec.~\ref{sec:weak_pairing_slow_switch}).
		The side panels show full heat-current transients $\Delta I_Q(t)$ in the same units [units $\Gamma U$] and, for comparison, the charge current transient $\Delta I_N(t)$ [units $\Gamma$]
		for selected triples of points marked in the main color-plot panels.
		All side panels share the same axis ticks.
		In the lower left panels, for $\Delta I_Q(t)$ the magenta curves lie on top of the black curves, hiding the latter, unlike Fig.~\ref{fig:currents_strong_pairing} where they are distinct.
	}
	\label{fig:currents_weak_pairing}
\end{figure*}
%%%%%%%%%%%%%%%%%%%%%%%%%%%%%%%%%%%%%%%%%%%%%%%%%%%%%%%
%\cleardoublepage 

%%%%%%%%%%%%%%%%%%%%%%%%%%%%%%%%%%%%%%%%%%%%%%%%%%%%%%%
\subsubsection{Fast switch.}
%%%%%%%%%%%%%%%%%%%%%%%%%%%%%%%%%%%%%%%%%%%%%%%%%%%%%%%
In Fig.~\ref{fig:amplitudes_strong_pairing} in column (a) we again first discuss the results
for large bias $|\mu| = \alpha = 2U$,
where the effects are well separated. The superconductor resonance is broadened due to the tenfold increase of $\alpha$ with respect to the weak-pairing regime just discussed,
but it can still be made out (longer grey dashed indicator).
It requires little further discussion
noting that the diagonal suppression (white) is trivial ($\epsilon_0=\epsilon$, no switch occurring).
The superconductor resonance persists in Fig.~\ref{fig:amplitudes_strong_pairing} as we step through distinct biasing conditions (a) $\to$ (a$'$) $\to$ (b), and,  as before, the vertical resonance disappears at low bias for (b$'$) $\to$ (c).

On this smooth background, the \emph{splitting} of Andreev levels
due to strong pairing $\alpha$ (Fig.~\ref{fig:thresholds})
now becomes apparent:
In Fig.~\ref{fig:amplitudes_strong_pairing}
there are sharp Andreev-state transitions (short grey dashed indicators),
which are only thermally broadened ($T \ll \alpha$),
counting four transitions marked~\marker{4} in (a), two marked~\marker{5} in (b) and strictly none in (c).
These Andreev transitions induced by the metal probe appear both in the initial and final gate voltage dependence of the fast switch.
As expected from Fig.~\ref{fig:invariants}, we see
for large bias in Fig.~\ref{fig:amplitudes_strong_pairing}(a)
that the left/lower two Andreev transitions coincide with the prominent CB resonances of the no-superconductor case
whereas the right/upper two are only weakly visible.
Upon lowering the bias the former CB-like Andreev resonances shift significantly
and one-by-one merge with their latter Andreev partners and disappear,
going from (a$'$) $\to$ (b) and from (b$'$) $\to$ (c), respectively.
For $a_c$ in particular in (b) between these two mergings there is a pronounced horizontal suppression
at the superconductor resonance.
Finally, at $\mu=0$ in (c) there are strictly no such resonances left.
This is intuitively expected from Fig.~\ref{fig:thresholds}(iii):
at $\mu=0$ and $\alpha \gg U$ the pairing dominates over interaction
and induces a gap on the hybridized dot of order $\alpha > U \gg T$,
destroying Coulomb blockade.

Finally, to identify more precisely which effects are due to interaction it is useful to compare with the $U=0$ limit.
Inspection of the fast-switch results for $U=0$ shown in App.~\ref{app:no_interaction}
reveals that cases (b)-(b$'$) and (c) of Fig.~\ref{fig:amplitudes_strong_pairing}
essentially survive for $U=0$ on top of a no-superconductor background which exhibits no Coulomb blockade. By contrast, in (a) and (a$'$) of Fig.~\ref{fig:amplitudes_strong_pairing}
the two-fold degeneracy of the Andreev transitions of the $U=0$ case is prominently broken
by the interaction.
See Ref.~\cite{Ortmanns22a} for further analytical comparison with the $U=0$ limit.

%%%%%%%%%%%%%%%%%%%%%%%%%%%%%%%%%%%%%%%%%%%%%%%%%%%%%%%
\subsubsection{Slow switch.}
%%%%%%%%%%%%%%%%%%%%%%%%%%%%%%%%%%%%%%%%%%%%%%%%%%%%%%%
In Fig.~\ref{fig:amplitudes_strong_pairing} inspection of the results 
for the slow switch shows that much of the comparison with the fast-switch results
from the weak-pairing case in Fig.~\ref{fig:amplitudes_weak_pairing}
carries over with two notable exceptions:
First, the regions where the slow-switch result survives while the fast switch result is suppressed
(dark red/green in the ratio plots) are much more prominent, see regions marked by~\marker{6}.

Second, there are now additional white panels in column~(c):
in these bias regimes the slow-switch amplitudes $a$ and $a_c$
are \emph{entirely thermally activated} --in addition to $a_p$--
remaining essentially zero at the considered low temperature until the first Andreev pair becomes accessible for intermediate bias as we move (c) $\to$ (b$'$).
This leads to similar non-monotonic $T$ dependence, see Sec.~\ref{sec:temperature}.

%%%%%%%%%%%%%%%%%%%%%%%%%%%%%%%%%%%%%%%%%%%%%%%%%%%%%%%
\subsection{Transient charge- and heat-current}\label{sec:reversal}
%%%%%%%%%%%%%%%%%%%%%%%%%%%%%%%%%%%%%%%%%%%%%%%%%%%%%%%

\begin{figure*}[t]
	\includegraphics[width=1.0\linewidth]{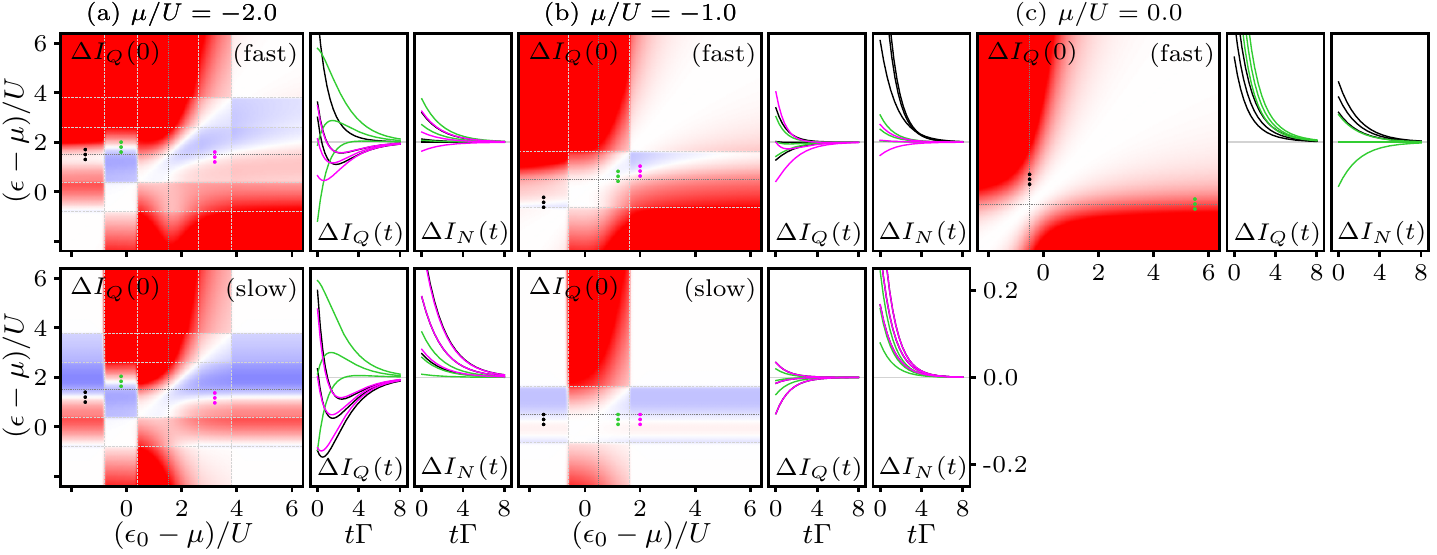}
	\caption{
		Strong pairing ($\alpha = 2.0\, U$): Transient heat current, see caption of Fig.~\ref{fig:currents_weak_pairing}.
		The ticks on all the side panels showing complete transients are again the same,
		but the value different from that used in Fig.~\ref{fig:currents_weak_pairing}.
	}
	\label{fig:currents_strong_pairing}
\end{figure*}

We now put the amplitudes together with the decay rates and outline the results for the directly accessible currents in Figs.~\ref{fig:currents_weak_pairing}-\ref{fig:currents_strong_pairing}.
For selected points (indicated in green, black and magenta) in the main panels
we show the full transient charge and heat currents for both the fast and slow switch.

For the single-exponential transient charge current $\Delta I_N(t)=I_N(t)-I_N(\infty)=a e^{-\gamma_c t}$
the decay rate $\gamma_c$ was already mapped out in Fig.~\ref{fig:invariants}.
As in Sec.~\ref{sec:no_pairing},
the sign of the initial transient  $\Delta I_N(0)=a$ decides whether the transient current goes with (red) or against (blue) the stationary charge flow.
If $\Delta I_N(0)=0$ (white) initially then it \emph{stays} zero due to the single exponential form,
$\Delta I_N(t)=0$ for $t \geq 0$.
The initial transient $\Delta I_N(0)=a$ was already plotted in Figs.~\ref{fig:amplitudes_weak_pairing}-\ref{fig:amplitudes_strong_pairing}
and the selected points considered here are also indicated there.
For weak pairing the side panels in Fig.~\ref{fig:currents_weak_pairing}
show full transients $\Delta I_N(t)$ for these points.
We see in particular at high bias that
as one steps through the superconductor resonance outside the final CB regime (black points)
the transient current is first comparable to the no-superconductor value (first point)
then suddenly suppressed (second point on the resonance) and then comparable again (the curve related to the third point coincides with the one of the first, hiding it).
This is clearly distinct from the no-superconductor case in Fig.~\ref{fig:amplitudes_no_pairing} (black points).
The same happens when stepping through the resonance inside the initial CB regime (green points).
As the bias is decreased this suppression at the superconductor resonance is lifted.
Finally, stepping through the resonant peak (magenta points) occurring on top of the zero background
the charge current amplitude stays small but, interestingly, reverses sign.
In Fig.~\ref{fig:currents_strong_pairing} we show that for strong pairing these effects
are qualitatively similar but more pronounced.

%%%%%%%%%%%%%%%%%%%%%%%%%%%%%%%%%%%%%%%%%%%%%%%%%%%%%%%
%%%%%%%%%%%%%%%%%%%%%%%%%%%%%%%%%%%%%%%%%%%%%%%%%%%%%%%

The double-exponential transient heat current $\Delta I_Q(t)=I_Q(t)-I_Q(\infty)=a_c e^{-\gamma_c t} + a_p e^{-\gamma_p t}$
showcases interesting deviations from the simple behavior of the charge current
 for both the fast and slow switch.
The initial transient value $\Delta I_Q(0)=a_c+a_p$ is plotted in the main panels of Figs.~\ref{fig:currents_weak_pairing}-\ref{fig:currents_strong_pairing}
together with full transients for the same selections of points.
Extreme deviations occur in particular for switching gate voltages
for which the initial transient is zero, $\Delta I_Q(0)=0$ 
[ignoring again trivial zeros at $\epsilon_0=\epsilon$].
In this case this does \emph{not} imply that $\Delta I_Q(t)$ stays zero
as discussed in Secs.~\ref{sec:heat_transient} and \ref{sec:no_pairing}.
Indeed, the side panels for weak pairing and more pronounced for strong pairing
exemplify switches for which the response is \emph{initially} zero,
then develops a transient, and finally \emph{again} goes to zero without sign change.
We refer to these as ``initially-zero transients''.
Close to such gate-voltage points
the direction of the transient heat current relative to the transient charge current
can even \emph{reverse} in time: whereas the two-particle contribution $a_p e^{-\gamma_p t}$ causes the transient heat current at $t=0$ to flow opposite to / along with the transient charge current,
due to $a_c e^{-\gamma_c t}$ it \emph{reverses} at intermediate times to flow along / opposite to it,
see Sec.~\ref{sec:nonmonotonicity} and Fig.~\ref{fig:nonmonotonicity}.
That the stationary charge and heat current can flow both in the same or in the opposite direction
due to competition of electron and hole contributions is of course well-known,
but here we see the nontrivial \emph{dynamics} of this competition.

We further observe for the fast switch in Figs.~\ref{fig:currents_weak_pairing}-\ref{fig:currents_strong_pairing} (top panels) that
as one steps through the superconductor resonance (black and green points)
the heat current transient is non-zero at the superconductor resonance --as noted in Sec.~\ref{sec:weak_pairing_fast_switch}--
in contrast to the charge current which vanishes there.
The sign of this two-particle energy current at the superconductor resonance can be strongly negative,
both in Figs.~\ref{fig:currents_weak_pairing} (blue along narrow lines)
and \ref{fig:currents_strong_pairing} (blue in extended regions).
Here the superconductor leads to a \emph{transient cooling} effect
relative to the dominating stationary heating of the metal.
As one steps through the resonance here (green points)
the transient heat current profile changes from positive decay (first point)
to an initially-zero transient (second point)
to negative decay (third point)
and then back (last two points).
For the slow switch in Figs.~\ref{fig:currents_weak_pairing}
and \ref{fig:currents_strong_pairing} (lower panels) the heat current vanishes along a horizontal line
close to the superconductor resonance.
Stepping through the resonance at high bias
at various positions (black, green and magenta points)
we likewise find initially-zero transients and reversals
which disappear with decreasing bias.

% LEAVE FOLLOWING EMPTY LINE HERE FOR mksubmit script !

\section{Discussion of features\label{sec:discussion}}

In this final section, we explain the main features identified in the above overview
using essentially the understanding of the \emph{stationary} problem.
We furthermore investigate the temperature dependence of the results.

\subsection{Gate-voltage dependence at low temperature}\label{sec:discussion_weak_pairing}

The behavior of the invariants~\eqref{eq:gammas}, observables~\eqref{eq:Ap_expectations}, and overlap~\eqref{eq:theta} 
 was already explained in terms of the stationary behavior of the system
and its dual system in Sec.~\ref{sec:invariants}.
We now in turn use this to explain the quite complex \emph{dynamic} response results of Sec.~\ref{sec:spectroscopy} for both types of switches using our duality-based formulas~\eqref{eq:charge_current}-\eqref{eq:heat_current}.
We focus on the dependence on the initial and final voltage defining the switch.
In Sec.~\ref{sec:discussion_weak_pairing}1-3
we first consider weak-pairing results for which the effects are clearly distinguishable and Andreev transitions which are CB-like are well defined.
We will indicate  by ``$\approx$'' those results that hold to a good approximation
in this regime.
% for $\alpha < U$.
These provide the underlying structure of the strong-pairing results in Sec.~\ref{sec:discussion_strong_pairing} on which we comment at the end.

\subsubsection{Transient charge current amplitude $a$ ($\alpha < U$)\label{sec:discussion_a}}

For the charge current amplitude
$a
= \gamma'_s
\big[ \brkt{A_0}_{\rho_0} -\brkt{A}_{z} \big]
$
[Eq.~\eqref{eq:a}]
the dependence on the final gate voltage ($\epsilon$)
at the superconducting resonance is dominated by the vanishing of the prefactor $\gamma'_s$,
the invariant accounting for electron-hole asymmetry relevant for transport:
$\gamma'_s=0$ if and only if $\epsilon=-U/2$ [Fig.~\ref{fig:invariants}(i)].
This explains the existence of the horizontal resonance in $a$ in Fig.~\ref{fig:amplitudes_weak_pairing}-\ref{fig:amplitudes_strong_pairing}
and its effect on $I_N$ in Fig.~\ref{fig:currents_weak_pairing}-\ref{fig:currents_strong_pairing},
independent of the type of switch.
Its occurrence is also independent of the bias,
whether $\epsilon=-U/2$ lies in the final CB regime ($-U < \epsilon-\mu < 0$) or not.
(Such dependencies would instead enter via the factor $\brkt{A_0}_{\rho_0}-\brkt{A}_{z}$ but are irrelevant since $\gamma'_s=0$.)

By contrast, the initial gate-voltage dependence ($\epsilon_0$) enters the charge amplitude $a$
only through $\brkt{A}_{\rho_0}=\theta \brkt{A_0}_{z_0}$ 
(since $\gamma'_s$ is $\epsilon_0$-independent).
Therefore, whether a vertical superconductor resonance exists or not now depends on the bias regime.
For large bias, $|\mu|>U/2$
%in the weak-pairing regime,
the initial polarization $\brkt{A_0}_{z_0}$ changes sign at $\epsilon_0=-U/2$
as seen in Fig.~\ref{fig:invariants}(i) (replace $\epsilon \to \epsilon_0$ in the plot of $\brkt{A}_{z}$).
We thus have a vertical superconductor resonance in $a$
at which the amplitude takes the nonzero value $a = -\gamma'_s \brkt{A}_{z}  $ (unless trivially $\epsilon=\epsilon_0$).
For low bias $|\mu|<U/2$, where interaction dominates over pairing,
the superconductor resonance instead lies inside the initial CB regime
throughout which $\brkt{A_0}_{z_0}=0$, see Fig.~\ref{fig:invariants}(i),
and therefore there is no special feature at $\epsilon_0=-U/2$.

We next explain how the signature of both the horizontal and vertical superconductor resonance
depends on the switch type [Eq.~\eqref{eq:Ap_initialize}]:

For large bias, we observed in Sec.~\ref{sec:weak_pairing} that these resonances have the signature of a peak / dip for the fast switch,
but an onset of a plateau for the slow switch.
This difference hence occurs when the superconductor (vertical / horizontal) resonance lies outside the (initial / final) CB regime and it is governed only by the excess-polarization factor $\brkt{A}_{\rho_0} -\brkt{A}_{z}= \theta \brkt{A_0}_{z_0} - \brkt{A}_{z}$
(it is irrelevant that $\gamma'_s=0$ precisely \emph{at} the horizontal resonance).
For switches between non-Coulomb-blockaded regions, the polarizations $\brkt{A_0}_{z_0}$ or $\brkt{A}_{z}$ in the vicinity of the superconductor resonance at
$\epsilon_0=-U/2$ resp. $\epsilon=-U/2$
 always takes nonzero values.
Moving $\epsilon_0$ or $\epsilon$  across the superconductor resonance changes either the sign of $\brkt{A_0}_{z_0}$ or $\brkt{A}_{z}$, leading
to $\brkt{A_0}_{z_0} \approx \brkt{A}_{z}$ on one side  and to $\brkt{A_0}_{z_0} \approx-\brkt{A}_{z}$ on the other side of the resonance.
For the slow switch ($\theta =1$), this leads to the cancellation of the two terms one one side, such that  $a=0$ and to the terms adding up to $a \neq 0$
on the other side. We thus have a transition to a plateau.
In contrast, for the fast switch
not only $\brkt{A_0}_{z_0}$ or $\brkt{A}_{z}$,
but also $\theta$ changes sign when moving across this resonance, see Fig.~\ref{fig:theta}.
This leads to having \emph{either} cancellation \emph{or} addition on \emph{both} sides of a resonance:
we thus have a dip or peak on a constant background.
This different behavior of the prefactor $\brkt{A_0}_{\rho_0} -\brkt{A}_{z}$
for switches between \emph{non-Coulomb-blockaded} even-parity regimes is intuitively expected, due to the possibility of exchanging Cooper pairs with the superconducting contact during a slow switch thereby retaining the initial polarization, see Sec.~\ref{sec:solution_duality}.

Finally, for low bias $|\mu| < U/2$ we observed in Sec.~\ref{sec:weak_pairing}
that only the horizontal resonance $\epsilon = -U/2$ survives.
It lies inside the initial CB regime as explained at the beginning of this section
(by the vanishing of $\gamma'_s$).
In this case we additionally noted in Fig.~\ref{fig:amplitudes_weak_pairing}(c)
that the charge amplitude $a$ changes sign across the resonance for the slow switch (resonant step)
but not for the fast switch (leaving a resonant dip).
This difference is explained in a similar way by the excess polarization factor
for which we have  $\brkt{A_0}_{\rho_0} -\brkt{A}_{z}=\theta$ here,
since $\brkt{A}_{z}=0$ at $\epsilon = -U/2$ for $|\mu| < U/2$
and $\brkt{A_0}_{z_0}=1$ (outside the initial CB regime).
Thus $a = \gamma'_s \theta$:
for the slow switch ($\theta=1$) the amplitude maps out the alternating electron-hole sign of $\gamma'_s$ around $\epsilon=-U/2$ giving the sign change
[$\brkt{A_0}_{z_0}$ is $\epsilon$-independent],
but for the fast switch the $\theta$ function~\eqref{eq:theta}
cancels this sign change producing a resonant dip.

\subsubsection{Transient heat current amplitude $a_c$  ($\alpha < U$)\label{sec:discussion_ac}}
\begin{figure*}[t]
	\includegraphics[width=1.0\linewidth]{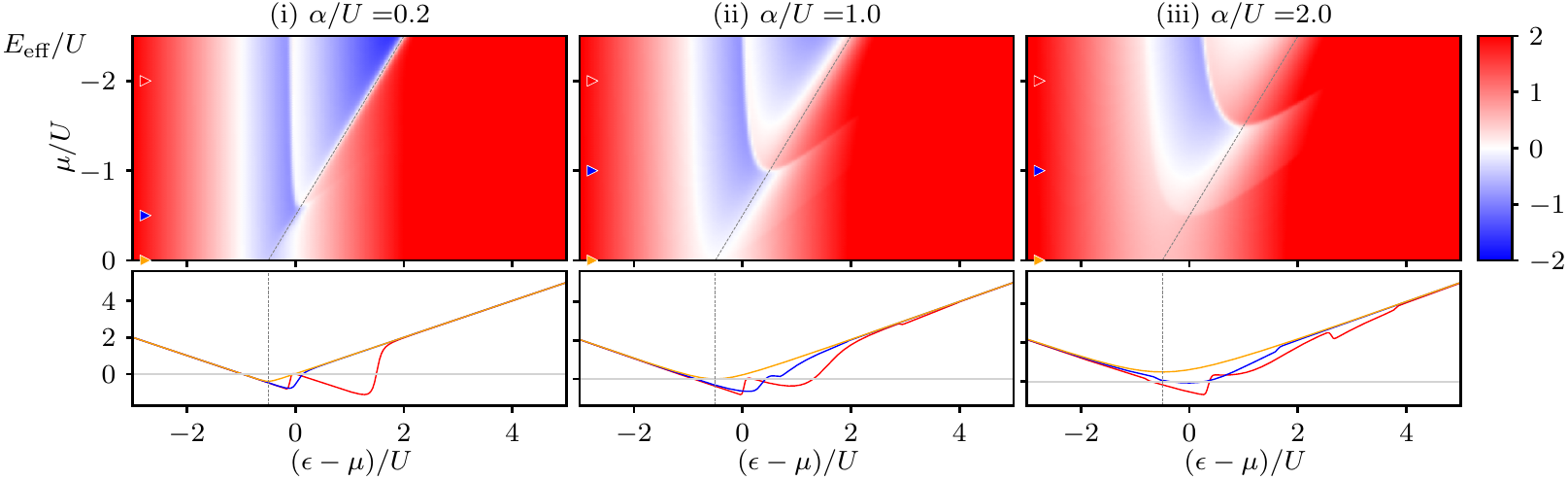}
	\caption{
		Effective energy~\eqref{eq:E_eff} in units of the interaction energy $U$
		as function of gate and bias voltage.
		Parameters and conventions are the same as in Fig.~\ref{fig:invariants}.
	}
	\label{fig:E_eff}
\end{figure*}

The heat current amplitudes $a_c$ and $a_p$ in Fig.~\ref{fig:amplitudes_weak_pairing}-\ref{fig:amplitudes_strong_pairing}
and $I_Q$ in Fig.~\ref{fig:currents_weak_pairing}-\ref{fig:currents_strong_pairing}
allow a similar detailed understanding.
For the amplitude
$a_c = E_\text{eff} \,  \gamma_c \big[ \brkt{A_0}_{\rho_0} -\brkt{A}_{z} \big]$
[Eq.~\eqref{eq:a_c}],
we additionally need to consider the effective energy
\begin{align}
	E_\text{eff} =\tfrac{1}{2} \big( \delta_\A - U \brkt{A}_{\bar{z}} \big)  -\mu \gamma'_s / \gamma_c
	\label{eq:E_eff}
	.
\end{align}
This energy prefactor is plotted in Fig.~\ref{fig:E_eff}
and can be understood as before from Fig.~\ref{fig:invariants}(i).
It is responsible for the appearance of additional zeros and sign changes in the amplitude $a_c$
as function of the final gate voltage $\epsilon$, depending on the bias and pairing.
The switch-specific explanations of the vertical resonance for $a_c$
follows from consideration of the initial gate voltage dependence
of $\brkt{A_0}_{\rho_0} -\brkt{A}_{z}$ alone,
exactly as for the charge amplitude $a$ in this case ($E_\text{eff} \, \gamma_c$ is $\epsilon_0$-independent).
Indeed, comparing with Eq.~\eqref{eq:a} we see that the ratios plotted in Eq.~\ref{fig:amplitudes_weak_pairing}-\ref{fig:amplitudes_strong_pairing}
are the same, determined only by the excess polarization ratio:
$a$(slow)/$a$(fast)=$a_c$(slow)/$a_c$(fast)
$= \big[ \brkt{A}_{z_0} -\brkt{A}_{z} \big]/\big[ \theta \brkt{A}_{z_0} -\brkt{A}_{z} \big]$.

%%%%%%%%%%%%%%%%%%%%%%%%%%%%%%%%%%%%%%%%%%%%%%%%%%%%%%%
%%%%%%%%%%%%%%%%%%%%%%%%%%%%%%%%%%%%%%%%%%%%%%%%%%%%%%%

To understand the existence of the horizontal resonance
we need to consider the final gate-voltage dependence of two factors
[$\gamma_c > 0$ plays no role in this, cf. Eq.~\eqref{eq:gamma_c_bounds}]:
In Fig.~\ref{fig:E_eff}(i)
we see that at high bias,
the effective energy $E_\text{eff}$
%always
has three zeros (red curve),
two at the CB-like Andreev transitions ($\epsilon -\mu \approx -U$ and $0$)
and one at the superconductor resonance $\epsilon \approx -U/2$. 
Thus for this case the heat current amplitude $a_c$ at the superconductor resonance $\epsilon=-U/2$ always shows a zero,
irrespective of the type of switch.
By contrast, for low bias, the zero at the superconductor resonance always disappears
from $E_\text{eff}$ (orange and blue curve), leaving only the two CB-like Andreev zeros in Fig.~\ref{fig:E_eff}(i).
Therefore, at low bias, the factor
$\brkt{A}_{\rho_0} -\brkt{A}_{z}=\theta$, analyzed at the end of Sec.~\ref{sec:discussion_a},
decides whether or not $a_c$ vanishes at the superconductor resonance $\epsilon=-U/2$.
We thus have $a_c \approx E_\text{eff} \, \theta$ which shows one zero for the fast switch,
but no zero for the slow switch.
This explains why \emph{only} for the fast switch we have 
$a_c \approx 0$ along the horizontal resonance $\epsilon=-U/2$ for all biases
as noted in Secs.~\ref{sec:weak_pairing_fast_switch}-\ref{sec:weak_pairing_slow_switch}.
This leaves the pure two-particle transient heat current $\Delta I_Q(t)= a_p e^{-\gamma_p t}$ whose amplitude we analyze next.

%%%%%%%%%%%%%%%%%%%%%%%%%%%%%%%%%%%%%%%%%%%%%%%%%%%%%%%%%%%%%%%%%%%%%%%%%%%%%%

\subsubsection{Transient heat current amplitude $a_p$  ($\alpha < U$)\label{sec:discussion_ap}}

The explanation of the behavior of amplitude
$a_p / \gamma_p  U = 
  \tfrac{1}{4}                      [  \brkt{p}_{\rho_0} -\brkt{p}_{z} ]
+ \tfrac{1}{2} \brkt{A}_{\bar{z}}  [ \brkt{A_0}_{\rho_0} -\brkt{A}_{z} ]
$
[Eq.~\eqref{eq:a_p}]
involves the competition of
two-particle contributions of excess parity and excess polarization,
which, interestingly, is
modulated by the stationary polarization $\brkt{A}_{\bar{z}}$
of the \emph{dual} system.

\emph{Superconducting resonance in $a_p$.}
We first discuss the superconductor resonance
focusing on the high bias case where it is independent of the switch type (magenta $a_p$ line cuts in Fig.~\ref{fig:amplitudes_weak_pairing}-\ref{fig:amplitudes_strong_pairing}).

At the horizontal resonance $\epsilon=-U/2$
we have $\brkt{p}_{z}=0=\brkt{A}_{z}$ 
and additionally $\brkt{A}_{\bar{z}}=0$ by Fig.~\ref{fig:invariants}(i)
such that 
$a_p / (\gamma_p  U) \approx \tfrac{1}{4} \brkt{p}_{z_0}$
maps out the parity of the \emph{actual} stationary system. As expected, as function of $\epsilon_0$, it exhibits signatures of Coulomb blockade:
$a_p / (\gamma_p  U)  = \pm \tfrac{1}{4} $ outside (inside) the initial CB regime.
As observed in Sec.~\ref{sec:weak_pairing_fast_switch}-\ref{sec:weak_pairing_slow_switch},
this holds for both the slow and fast switch.
The reason is that the factor $\brkt{A}_{\bar{z}}=0$ cancels out the switch-dependent contribution $\brkt{A}_{\rho_0}=\theta \brkt{A_0}_{z_0}$.
Since we found that along this horizontal resonance $a_c = 0$,
the Coulomb blockade enters the transient heat current only through
the pure two-particle contribution, $\Delta I_Q(t) = a_p e^{-\gamma_p t}$.

At the vertical resonance, $\epsilon_0=-U/2$,
we have $\brkt{p}_{z_0}=0=\brkt{A_0}_{z_0}$ for $|\mu| > U$ in Fig.~\ref{fig:invariants}(i), 
such that again the switch-dependence then drops out. We get $a_p / (\gamma_p  U)
=
- [ \tfrac{1}{4} \brkt{p}_{z} + \tfrac{1}{2} \brkt{A}_{\bar{z}} \brkt{A}_{z} ]
=
\tfrac{1}{4} \brkt{p}_{\bar{z}}$,
where the last equality is a nontrivial relation between stationary observables of the actual and the dual model, holding for any set of parameters (Eq.(44) of Ref.~\cite{Ortmanns22a}).
Thus, along the vertical superconductor resonance $a_p / \gamma_p  U = \tfrac{1}{4} \brkt{p}_{\bar{z}}$
is \emph{constant} as function of $\epsilon$.
This seems to defy simple explanation in terms of the repulsive interaction of the actual system or by the promotion of even-parity states by the pairing $\alpha$ since it already holds for \emph{weak} pairing relative to interaction (and continues to hold for strong pairing).
This value of $a_p$ instead maps out the stationary parity
of the \emph{dual} system, which 
by Fig.~\ref{fig:invariants}(i) equals $\brkt{p}_{\bar{z}}=1$ everywhere
[except trivially at the crossing $\epsilon = \epsilon_0=-U/2$ where we don't switch].
Clearly, this observable \emph{cannot} show any signature of repulsive Coulomb blockade as explained in Sec.~\ref{sec:invariants}.
The surprising \emph{lack} of an effect in $a_p$ along the vertical resonance in the final CB regime, noted in Sec.~\ref{sec:weak_pairing_fast_switch}-\ref{sec:weak_pairing_slow_switch},
thus receives a very simple physical explanation by duality.
In the observable transient heat current $I_Q(t)$ the repulsive Coulomb blockade effect is instead expressed
through $a_c$ which sharply changes as function of $\epsilon$ along the vertical resonance,
vanishing in the final CB regime ($-U < \epsilon -\mu < 0$).
In this regime the resulting pure two-particle transient heat current $\Delta I_Q(t) = a_p e^{-\gamma_p t}$ nevertheless probes exclusively the parity of the \emph{attractive} interaction in the \emph{dual} model.

\emph{Plateaus of $a_p$.}
It remains to explain how the plateau values of $a_p$ arise from the competition between parity and polarization.
The excess parity $ \brkt{p}_{\rho_0} -\brkt{p}_{z}$ is easily understood
and does not depend on the type of switch [Eq.~\eqref{eq:Ap_initialize}].
At low bias $|\mu| < U/2$ and weak pairing $\alpha<U$, we see in Fig.~\ref{fig:invariants}(i)
that each parity takes the value $1$ outside the CB regime and $-1$ inside.
We thus have $ \brkt{p}_{\rho_0} -\brkt{p}_{z}= \pm 2$
if $\epsilon$ ($\epsilon_0$) lies inside the CB regime
and $\epsilon_0$ ($\epsilon$) lies outside of it.
For all other switches $\brkt{p}_{\rho_0} -\brkt{p}_{z}=0$ at low bias.
At high bias $|\mu| \geq U/2$
this result still holds with one exception:
 the parity can vanish at the superconductor resonance now lying outside the CB regime, as discussed above.
The excess polarization term $\brkt{A}_{\bar{z}}  [ \brkt{A_0}_{\rho_0} -\brkt{A}_{z} ]$
now brings in a dependence on the switch type:

\emph{Slow switch.}
At low bias for the slow switch ($\theta=1$)
the dual polarization equals $\brkt{A}_{\bar{z}} \approx 1$ for all $\epsilon$ by Fig.~\ref{fig:invariants}(i).
We also see that
$\brkt{A_0}_{\rho_0} -\brkt{A}_{z}\approx \pm 1 \approx \tfrac{1}{2} [ \brkt{p}_{\rho_0} -\brkt{p}_{z} ]$
if $\epsilon$ ($\epsilon_0$) lies inside the CB regime
and $\epsilon_0$ ($\epsilon$) lies outside it.
Thus, the excess polarization and excess parity cancel out
to give $a_p\approx0$ for \emph{all} gate voltages---as long as the temperature is low.
The complete suppression of $a_p$ for the slow switch
observed in Sec~\ref{sec:weak_pairing_slow_switch}
thus arises from this competition of polarization and parity
and leads to thermally activated behavior illustrated in Sec.~\ref{sec:temperature}.

At high bias, inspection of Fig.~\ref{fig:invariants}(i) shows that two things change
in the gate-voltage plane:
we can have dual polarization $\brkt{A}_{\bar{z}}\approx - 1$ and simultaneously $\brkt{A}_{z}\approx+1$
in the horizontal strip $\mu < \epsilon < -U/2$ in Fig.~\ref{fig:amplitudes_weak_pairing}
and we can have $\brkt{A_0}_{z_0}\approx+1$ in the vertical strip $\mu < \epsilon_0 < -U/2$.
Inside these regimes a nonzero value of $a_p$ is possible
which turns out to always be $a_p/\gamma_p U \approx 1$:
we either have $\brkt{A}_{\bar{z}}  [ \brkt{A_0}_{\rho_0} -\brkt{A}_{z} ] \approx 2$
and no excess parity $\brkt{p}_{\rho_0} -\brkt{p}_{z}\approx0$~\footnote
	{Cases to consider:
	$\brkt{A_0}_{z_0}\approx-1$,
	$\brkt{A}_{z}\approx 1$,
	$\brkt{A}_{\bar{z}}\approx1$
	and
	$\brkt{A_0}_{z_0}\approx 1$,
	$\brkt{A}_{z}\approx -1$,
	$\brkt{A}_{\bar{z}}\approx1$.
}
or we have $\brkt{A}_{\bar{z}}  [ \brkt{A_0}_{\rho_0} -\brkt{A}_{z} ] \approx 1$
and excess parity $\brkt{p}_{\rho_0} -\brkt{p}_{z}\approx2$~\footnote
 	{Cases to consider:
 	 $\brkt{A_0}_{z_0}\approx 1$,
 	 $\brkt{A}_{z}\approx 0$,
 	 $\brkt{A}_{\bar{z}}\approx1$.
 	}.
In the remaining two cases in these regimes one still obtains $a_p\approx0$~\footnote
	{Cases to consider:
	$\brkt{A_0}_{z_0}\approx 1$,
	$\brkt{A}_{z}\approx 1$,
	$\brkt{A}_{\bar{z}}\approx-1$
	and
	$\brkt{A_0}_{z_0}\approx 0$,
	$\brkt{A}_{z}\approx 1$,
	$\brkt{A}_{\bar{z}}\approx-1$.
	}.
This explains how the suppression of $a_p$ just discussed
is lifted when the bias voltage makes positive values of the polarization and negative values of the \emph{dual} polarization accessible.
Outside these strips the situation is the same as for low bias and the cancellation to $a_p\approx0$ persists.

\emph{Fast switch.}
Finally, it remains to explain which modifications to the above occur for the fast switch
reminding that the excess parity $\brkt{p}_{\rho_0} -\brkt{p}_{z}$ is the same
	for both switch types.
The switch-dependent excess polarization term is only modified with respect to the slow-switch result
in regimes where we have $\theta \approx - 1$.
At low bias, 
where $\brkt{A}_{\bar{z}} \approx 1$, these modifications occur in just two cases~\footnote
	{In one further case a deviation from the slow switch \emph{could} occur but it does not:
	(iii)
	For $\epsilon > -U/2 > \mu$ outside CB regime [$ \brkt{A}_{z} \approx 1$, $\brkt{A}_{\bar{z}}=1$]
	and $\epsilon_0$ inside the CB regime [$\brkt{A_0}_{z_0} \approx 0$]
	the switch dependence cancels out:
	$\brkt{A_0}_{\rho_0} -\brkt{A}_{z} = -\brkt{A}_{z} \approx 1$.}:
First, if $\epsilon$ lies inside the CB regime
and $\epsilon_0$ lies outside it,
then
$\brkt{A_0}_{\rho_0} -\brkt{A}_{z} = -\theta$
and
$\brkt{p}_{\rho_0} -\brkt{p}_{z} = 2$ using Fig.~\ref{fig:invariants}(i).
Thus $a_p/(U\gamma_p)= (1-\theta)/2$ which is a unit step located at $\epsilon=-U/2$.
This is the signature of the superconductor resonance
shifting relative to the final CB regime ($-U < \epsilon-\mu < 0$) with varying the bias as observed in Sec.~\ref{sec:weak_pairing_fast_switch}.
Second, if $\epsilon$ and $\epsilon_0$ lie on opposite sides of the CB regime
then
$\brkt{A_0}_{\rho_0} -\brkt{A}_{z} = - \brkt{A_0}_{z_0} -\brkt{A}_{z} = 2$
while $\brkt{p}_{\rho_0} -\brkt{p}_{z}=0$.
Thus $a_p/(U\gamma_p)=2$, implying that the plateau reached after the step in (i)
continues outside the CB regime as observed in Sec.~\ref{sec:weak_pairing_fast_switch}. 
Finally, for high bias we note that the superconducting resonance does not lead to the onset of plateaus for $a_p$ for the fast switch, see Sec.~\ref{sec:fast_switch}.

%%%%%%%%%%%%%%%%%%%%%%%%%%%%%%%%%%%%%%%%%%%%%%%%%%%%%%%
\begin{figure*}[t]
	\includegraphics[width=1.0\linewidth]{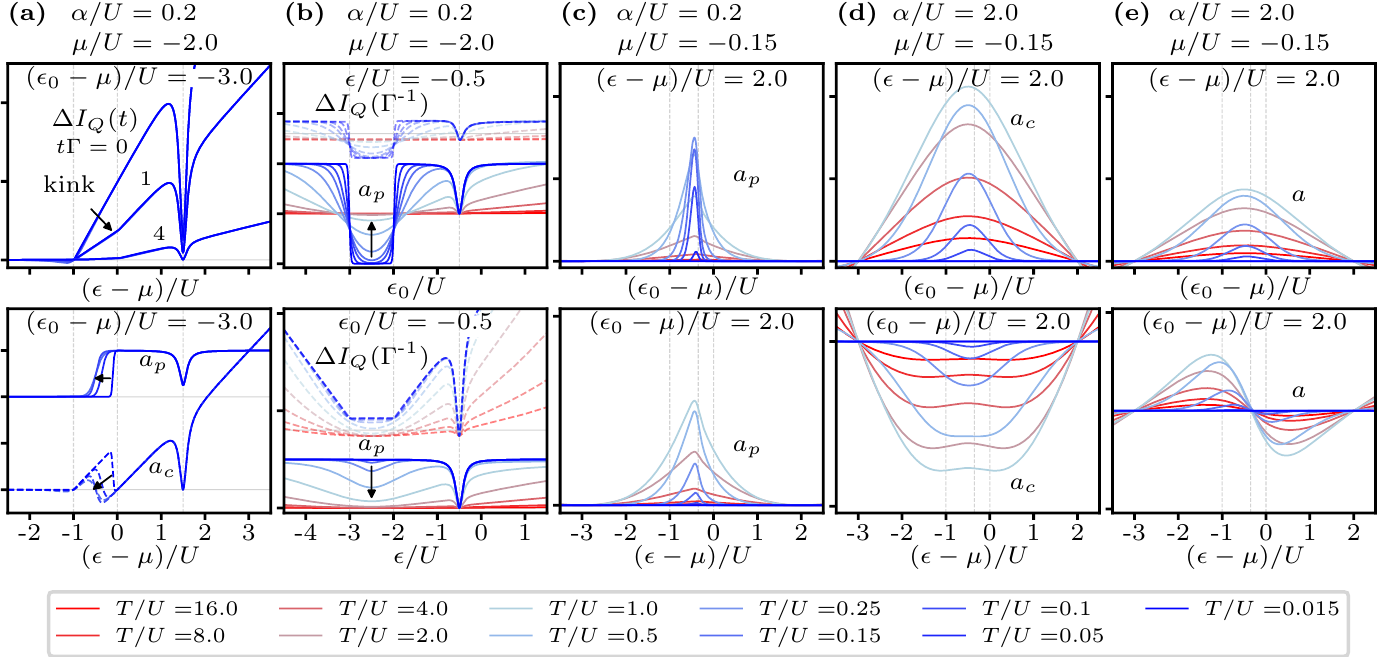}
	\caption{
		Temperature dependence of amplitude
		$a$ in units of $\Gamma$ and $a_c$, $a_p$, and $\Delta I_Q$ in units of $\Gamma \, U$
		for (a) fast switch,
		(b) slow and fast switch (same),
		and (c)-(e) slow switch.
		In all figures the tick-mark distance on the vertical axis equals $0.25$ and
		zero is indicated by a horizontal grey line.
		In (a) and (b) the dashed curves are offset for clarity (see text).
		The left-most two vertical grey lines indicate the Coulomb blockade regime $-U < \epsilon -\mu < 0$
		and the right vertical grey line marks the position of superconducting resonance $\epsilon = -U/2$ resp. $\epsilon_0=-U/2$.
	}
	\label{fig:temperature}
\end{figure*}
%%%%%%%%%%%%%%%%%%%%%%%%%%%%%%%%%%%%%%%%%%%%%%%%%%%%%%%

\subsubsection{Strong pairing  ($\alpha > U$)\label{sec:discussion_strong_pairing}}

The explanation of the gate-voltage dependence for weak pairing $\alpha < U$
focused on the superconductor resonance, the only pronounced feature.
As observed in Sec.~\ref{sec:strong_pairing} there are two
main changes for strong pairing $\alpha > U$:
Quantitatively,
the superconductor resonance gets broadened, smoothening the features explained earlier.
Qualitatively new is the onset of the Andreev resonances (\marker{4}, \marker{5} in Fig.~\ref{fig:amplitudes_strong_pairing}).
Their appearance at low temperature is directly understood from their appearance in the duality invariants and stationary observables in Fig.~\ref{fig:invariants}(iii) as sharp steps which was already explained in Sec.~\ref{sec:invariants}.

One additional feature remains to be explained.
For the slow switch at low bias, we observed \emph{only} for $\alpha > U$
a complete suppression of the amplitudes $a$ and $a_c$.
Since these are proportional to the excess polarization $\brkt{A_0}_{z_0} -\brkt{A}_{z}$
this follows from the lifting of the Coulomb blockade at low bias
giving equal stationary polarizations,
$\brkt{A_0}_{z_0} =\brkt{A}_{z}=-1$ for all $\epsilon_0$, $\epsilon$,
which cancel if $\theta =1$.
This cancellation can only be undone by thermal activation.
(By contrast, $a_p$ is suppressed by cancellation of two-particle contributions and occurs already for small pairing $\alpha < U$, see end of previous section).
For the fast switch $a$ and $a_c$ are likewise suppressed at low bias except in those regions where the switch-dependent parameter $\theta < 0$ spoils cancellation of terms
leading to nonzero $a$ and $a_c$.
Their signs equal the sign of $\gamma'_s$ and $E_\text{eff}>0$, respectively.

% LEAVE FOLLOWING EMPTY LINE HERE FOR mksubmit script !

\subsection{Temperature effects}\label{sec:temperature}

The weak-coupling results~\eqref{eq:charge_current}, \eqref{eq:heat_current} and \eqref{eq:Ap_initialize}
describe the transient response including the competition of pairing and thermal fluctuations for
$T, \alpha \gg \Gamma$
focusing on $T \ll U$.
With increasing temperature it becomes more difficult to disentangle the variety of features that we were able to discern so far and we highlight some interesting aspects of the $T$-dependence. One should remember that our infinite-gap approximation  results (no quasiparticles) are applicable to real, finite-gap superconductors only at temperatures $T \ll \Delta$ [Sec.~\ref{sec:model}] (no quasiparticles).

\subsubsection{Robustness against thermal fluctuations}

An immediate question concerns the robustness of the discussed features against temperature increase.
The features due to non-Coulomb-blockade type of Andreev transitions
are directly sensitive to thermal smearing
as seen in Fig.~\ref{fig:temperature}.
Like any other transition induced by the metal the relevant thermal energy scale is $4 T$, the width of the Fermi-function step.

By contrast, Fig.~\ref{fig:temperature}(a)-(b) illustrates
that the various signatures of the superconductor resonance
for both the fast and slow switch (peaks / dips or plateau-steps, respectively)
do not change their  width as function of gate-voltage as $T$ is increased; the width is instead  $\propto \alpha$.
Throughout all panels, curves computed at the same temperature have the same color.
The line cuts plotted  in Fig.~\ref{fig:temperature}(a) are taken vertically through $\epsilon_0-\mu=-3U$ in the fast switch plots of Fig.~\ref{fig:amplitudes_weak_pairing} and in the bottom panel show that
the dip at $\epsilon=-U/2$ in $a_c$ and $a_p$ (indicated by the marker \marker{1}) are clearly unaffected by temperature. Therefore, at this position the same holds for the resulting transient heat current $\Delta I_Q(t)$ at small, intermediate and large times ($t\gamma_p=0,1,4$). In particular, in the plot of $\Delta I_Q(t)$ in the upper panel, the four curves for different temperatures lie almost exactly on top of each other.

Fig.~\ref{fig:temperature}(b) shows the $T$ dependence of amplitude $a_p$
for high bias $|\mu| > U/2$.
It is taken along the horizontal superconductor resonance (cut through marker \marker{1} in Fig.~\ref{fig:amplitudes_weak_pairing}(a), top panel in Fig.~\ref{fig:temperature}(b))
and along the vertical one [cut through marker \marker{2} in Fig.~\ref{fig:amplitudes_weak_pairing}(a), bottom panel in Fig.~\ref{fig:temperature}(b)].
These are the same for both the fast and slow switch.
As mentioned in Sec.~\ref{sec:discussion_weak_pairing},
this essentially maps out the parity of the actual and dual system, respectively.
Indeed, along the horizontal resonance the interaction $U > \alpha$ causes a sign inversion of the parity $a_p/(\gamma_pU)=\brkt{p}_{z}/4$ in the CB regime, which is lifted with increasing $T$.
This determines the transient heat current $\Delta I_Q(t)$ at all times since at the superconductor resonance, $\epsilon=-U/2$, $a_c$ is suppressed for all $\epsilon_0$
as illustrated in the top panel of Fig.~\ref{fig:temperature}(b) (offset dashed curves for $t\gamma_p = 1$).
As mentioned earlier, along the vertical resonance there is surprisingly no signature of the interaction in $a_p/(\gamma_pU)=\brkt{p}_{\bar{z}}/4=1/4$ [vertical red line in Fig.~\ref{fig:amplitudes_weak_pairing}(a)]
but such a feature seems to \emph{develop with increasing} $T$.
This is immediately understood by duality:
the effect of the attractive interaction $\bar{U}=-U$ in the dual model, leading to a constant $\brkt{p}_{\bar{z}}=1$ (see Fig.~\ref{fig:invariants}), is suppressed with increasing temperature.
The signature of the actual interaction $U > \alpha$ on the transient heat current $\Delta I_Q(t)$
is instead imprinted by $a_c$ (not shown).
When it is combined to obtain $\Delta I_Q(t)$ (offset dashed curves bottom panel of Fig.~\ref{fig:temperature}(b))
its temperature dependence cancels out features developing in $a_p$.

Finally, we have verified that at higher temperatures ($T/U \approx 1/4$)
the spectroscopy plots in Fig.~\ref{fig:currents_weak_pairing}-\ref{fig:currents_strong_pairing}
still exhibits pronounced lines of gate-voltage points around the superconductor resonance  (e.g., the magenta points)
where $\Delta I_Q(0)=a_c + a_p$ vanishes nontrivially by cancellation (white line segments). 
Here the full transients remain strongly non-monotonic as in Sec.~\ref{sec:reversal}.
Interestingly, this effect now coexists with the different nonmonotonicity effects which occur also \emph{without} the superconductor at this high temperature, see App.~\ref{app:reversal}.
%The same holds for Fig.~\ref{fig:currents_weak_pairing}.

\subsubsection{Entropic resonance shifts linear in temperature}\label{sec:andreev}

The temperature evolution of the line cuts of amplitudes $a_c$ and $a_p$  in Fig.~\ref{fig:temperature}(a) displays another interesting thermal effect (lower panel):
Already for small increasing temperatures, the positions of the features in the CB regime
deviate noticeably from the energy thresholds. This shift is linear in $T$.
All other features in the plot remain unaffected up to $T \lesssim 0.15 \, U$ (first 4 deep blue line cuts).
This shift is a well-known general phenomenon~\cite{Bonet02,Deshmukh02}
for \emph{stationary} transport through weakly coupled systems with transition rates of varying magnitudes~\cite{Golovach2004Jun,Romeike2006May}
and indeed occur in the stationary results in Fig.~\ref{fig:stationary_current}(i) at $|\mu| < U/2$ when increasing $T$ (not shown).
They call for extra care in both experimental / theoretical analysis since
the naive direct identification of measured / computed data with energy-thresholds can be misleading,
resulting in apparent inconsistencies when varying temperature~\cite{Deshmukh02}. These shifts can be attributed to the different degeneracy of the eigenstates of the proximized dot.
Indeed, such shifts have recently been exploited to measure the entropy associated with the level degeneracy in quantum dots~\cite{Hartman2018Nov,Child2022}.

Our results show that such shifts also occur in the resonant parameter dependence of \emph{time-dependent} response of transport to a switch
and turn out to be present even in the no-superconductor limit.
They have received little attention so far,
whereas their effect on slowly driven transport dynamics has been considered~\cite{Calvo2012Dec,Riwar2013May}.
Our duality-based formulas~\eqref{eq:state}-\eqref{eq:heat_current}
rationalize their occurrence since they express the transient response in terms of stationary observables of the system \emph{and the dual} system
which we know exhibit such shifts.
What is quite subtle here is that the significant shifts for
$a_c$ and $a_p$ occurring in the first 4 curves  (lower panel)
\emph{cancel out} to produce the \emph{same} transient:
In the upper panel of Fig.~\ref{fig:temperature}(a) the transients $\Delta I_Q(t)$ collapse to essentially the same curve on short, intermediate and long times.
This cancellation of the temperature-dependent contributions to $a_c$ and $a_p$ 
occurs in the CB region where the charge rate $\gamma_c$ saturates the bound $\gamma_p$. The only remainder of the sharp steps in $a_c,a_p$ in the transient heat current is a kink at the onset of the CB regime which becomes more pronounced with time and is visible for both weak and strong pairing.

\subsubsection{Activation by thermal fluctuations}

Finally, the question of thermal activation came up in Sec.~\ref{sec:spectroscopy}:
for low bias, the slow-switch amplitudes can be entirely suppressed at low temperature $T \ll U$.
In Fig.~\ref{fig:temperature}(c)-(e) we show that temperature indeed activates these amplitudes.
Further increase of $T \gtrapprox U$ again suppresses the activated features by thermal smearing as usual.
This gives distinct nonmonotonic dependencies on $T$ reflecting the difference of the effects:
$a$ and $a_c$ are suppressed at low $T$ only for strong pairing [Sec.~\ref{sec:strong_pairing}]. This is expected intuitively due to the gap opening up  $\propto \alpha$
(strongly affecting all the invariants in Fig.~\ref{fig:invariants} that determine these amplitudes).
By contrast, $a_p$ is suppressed at low $T$ already for small pairing $\alpha < U$ [Sec.~\ref{sec:weak_pairing}]
due to a subtle cancellation of two-particle contributions [end of Sec.~\ref{sec:discussion_weak_pairing}]
and remains suppressed for $\alpha > U$.

In contrast to the slow switch, the decay after the fast switch does not require thermal activation. This is due to the fact that fast switches lead to a non-vanishing excess polarization essentially whenever at the initial gate voltage the dot is
empty and at the final gate voltage it is doubly occupied in the stationary state or vice versa. In this case, the initial state has a $|+)$ component, the decay of which results in a transient heat current with a two-particle component for $\alpha \neq 0$.
For these switches, the excess parity is always zero and cannot cancel out the excess polarization.
Therefore the fast-switch amplitudes have nonzero contributions in low-temperature regimes where the slow-switch amplitudes are fully suppressed.

% LEAVE FOLLOWING EMPTY LINE HERE FOR mksubmit script !
\section{Summary}\label{sec:summary}

We analyzed transient charge and heat transport spectroscopy
where a weakly coupled metal probes an interacting quantum dot which is proximized by a large-gap superconductor.
We focused on initial states which are mixtures of Andreev states,
and showed how these can be prepared in two ways, using either a fast or a slow gate-voltage switch ($t_0 \ll \alpha^{-1}$ resp. $t_0 \gg \alpha^{-1}$). These define distinct experiments and we exhaustively investigated the ensuing transient charge and heat transport
on the background of their stationary finite-bias currents.

For weak pairing relative to the interaction ($\alpha < U$)
the superconductor \emph{pair resonance} is the main feature
occurring along side Coulomb-blockade transient responses.
It appears when switching the quantum dot's symmetry point either towards or away from 
alignment with the superconductor
%($\epsilon+ U/2 = \mu_\S$ or $\epsilon_0+ U/2 = \mu_\S$),
irrespective of the alignment with the metal probe.
% ($\mu$).
We found that at this resonance the transient heat current
can be dominated by its \emph{two-particle} amplitude
% ($a_p$)
reflecting the importance of electron pairs.
The pairing leads to pronounced signatures in the charge and heat current decay
by generating electron and hole currents whose contributions can cancel out or add up.
 
For strong pairing ($\alpha > U$)
additional thermally-sharp features, induced by the metal probe, appear on the smooth background of the now broadened pair resonance.
These correspond to transitions between the Andreev states
and show up pairwise with energies clearly split by the repulsive interaction,
leading to a complex bias voltage dependence of the transient response.
The superconductor furthermore induces non-trivial double-exponential
heat-current decay profiles already at low temperature relative to the interaction ($T \ll U$).
Remarkably, these can feature a local maximum and also a preliminary crossing of the stationary value already at short times.
This effect is clearly tied to the pairing induced by the superconductor and its interplay with the interaction and the transport bias.
It is distinct from similar effects which occur \emph{without} a superconductor
at higher temperature ($T \sim U$) which we found here and that went unnoticed in Ref.~\cite{Schulenborg2016Feb}.

Temperature also activates the transient behavior for the slow switch response
% (amplitudes $a$, $a_c$)
for strong pairing as expected by the induced gap.
Interestingly, the two-particle heat amplitude
% ($a_p$)
already shows such activation for weak pairing
leading to similar non-monotonic $T$ dependence.
Finally, we found that significant shifts of resonant features with a strong linear temperature dependence~\cite{Bonet02,Golovach2004Jun,Romeike2006May,Deshmukh02,Hartman2018Nov,Child2022} can also occur in the two amplitudes of \emph{transient} response,
but these features remarkably cancel out in the total observable heat current.

With the continuing progress in detecting the charge and energy of individual electrons in a time-resolved manner~\cite{Fletcher2013Nov,Fletcher2019Nov,Ubbelohde2015Jan} in semiconductor nanostructures,
our results motivate extension of these experimental works to hybrid  superconducting systems.

% LEAVE FOLLOWING EMPTY LINE HERE FOR mksubmit script !
\acknowledgments

We are grateful for helpful discussions with Attila Geresdi. L.\,C.\,O. acknowledges support by the Deutsche Forschungsgemeinschaft (RTG~1995).
J.S. acknowledges financial support from the Swedish Vetenskapsr{\aa}det (project number 2018-05061)
and the Knut and Alice Wallenberg Foundation through the Fellowship program.

% LEAVE FOLLOWING EMPTY LINE HERE FOR mksubmit script !
\appendix
\section{Initial condition}\label{app:initial_condition}

Eq.~\eqref{eq:Ap_initialize} is obtained by expanding the initial state, expressed in the form of Eq.~\eqref{eq:bloch},
as follows:
For both switches, the change of the gate voltage $\delta_0 \to \delta$ leads to a change of the basis
from $\{\Sket{\one}, \Sket{A_0},\Sket{p} \} \to  \{\Sket{\one}, \Sket{A},\Sket{p} \}$
where $A_0$ and $A$ are the \emph{different} polarization operators at gate voltage $\delta$ and $\delta_0$,
respectively. This amounts to
\begin{align}
	\Sket{z_0}       & = \tfrac{1}{4} \Sket{\one} + \brkt{A_0}_{z_0} \tfrac{1}{2} \Sket{A_0} + \brkt{p}_{z_0} \tfrac{1}{4} \Sket{p}
	\notag\\
	\to \Sket{\rho_0} & = \tfrac{1}{4} \Sket{\one} + \theta \brkt{A_0}_{z_0} \tfrac{1}{2} \Sket{A}   + \brkt{p}_{z_0} \tfrac{1}{4} \Sket{p}
	\label{eq:rho0_fast_A}
	.	
\end{align}
for the fast switch~\eqref{eq:rho0_fast}, and yields the result for the slow switch~\eqref{eq:rho0_slow} when setting $\theta=1$.
Here we have used $\tfrac{1}{2} \Sbraket{A}{A_0} = \tfrac{1}{2}\sum_{\tau\tau'} \tau \tau' \Sbraket{\tau}{\tau'_0}
= \tfrac{1}{2} \sum_{\tau\tau'} \tau \tau' |\braket{\tau}{\tau'_0}|^2=\theta$.

% LEAVE FOLLOWING EMPTY LINE HERE FOR mksubmit script !
\section{No interaction ($U=0$)}\label{app:no_interaction}

In Fig.~\ref{fig:no_interaction}
we show how the strong pairing results in Fig.~\ref{fig:amplitudes_strong_pairing}
are modified when setting $U = 0$.
To facilitate comparison with Fig.~\ref{fig:amplitudes_weak_pairing} we continue using $U$ of the interacting case
as a \emph{reference} energy for the case $U=0$ to normalize $\epsilon_0-\mu$, $\epsilon-\mu$, $\mu$ and $\alpha$ in the plot.
The weak pairing results for $U=0$ are similar (not shown):
the superconductor resonance seen in Fig.~\ref{fig:amplitudes_weak_pairing} is sharpened and while Andreev transitions are suppressed.

% LEAVE FOLLOWING EMPTY LINE HERE FOR mksubmit script !
\section{Transient heat current reversal ($T\sim U$)}\label{app:reversal}

\begin{figure*}[t]
	\includegraphics[width=1.0\linewidth]{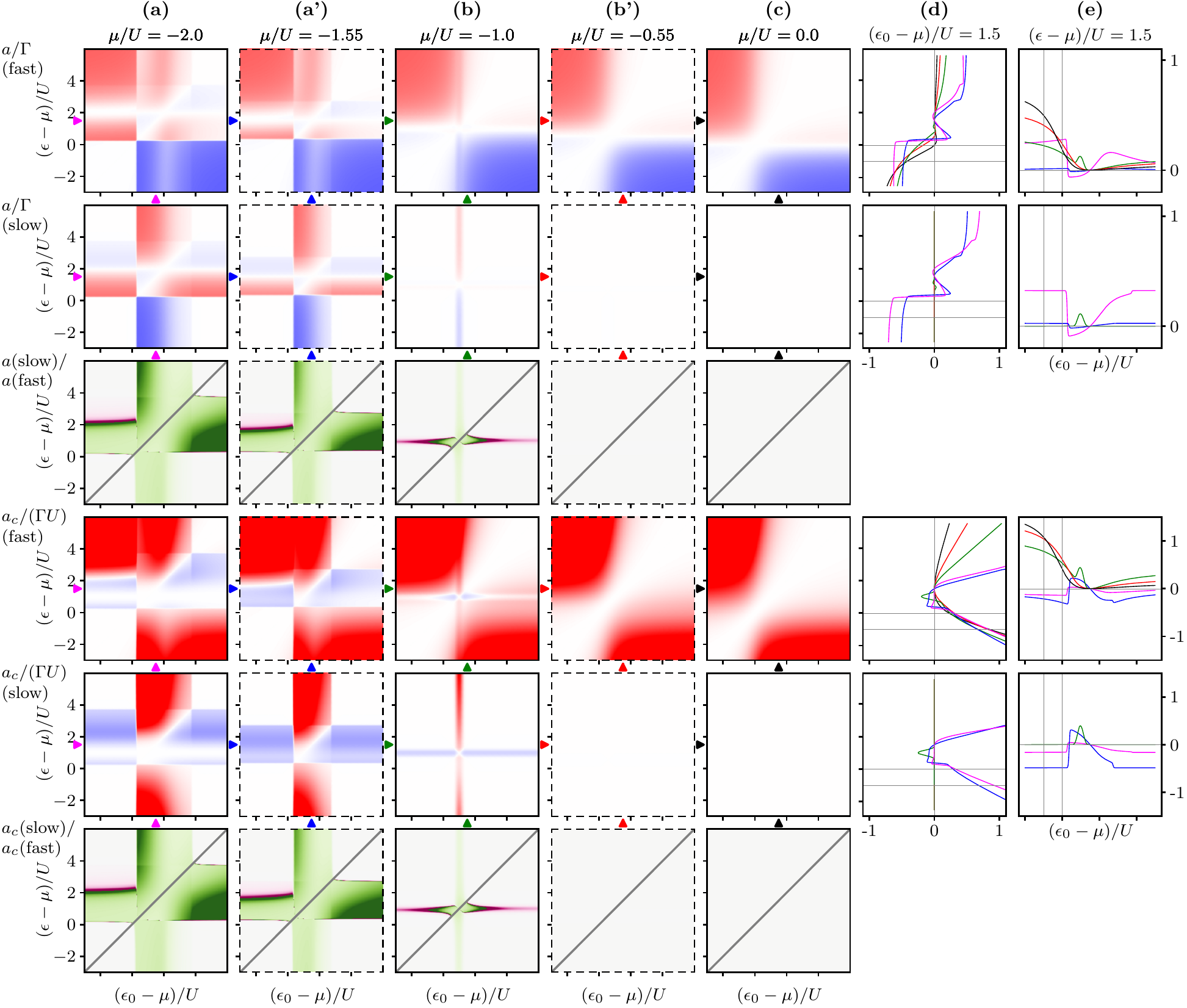}
	\caption{
		No interaction ($U=0$):
		Charge (heat) amplitudes $a$ ($a_c$) in units of coupling $\gamma_p=\Gamma$ ($\times$ interaction energy $U$),
		see caption Fig.~\ref{fig:amplitudes_weak_pairing} and main text for description of the layout.
	}
	\label{fig:no_interaction}
\end{figure*}

\begin{figure}[t]
	\includegraphics[width=1.0\linewidth]{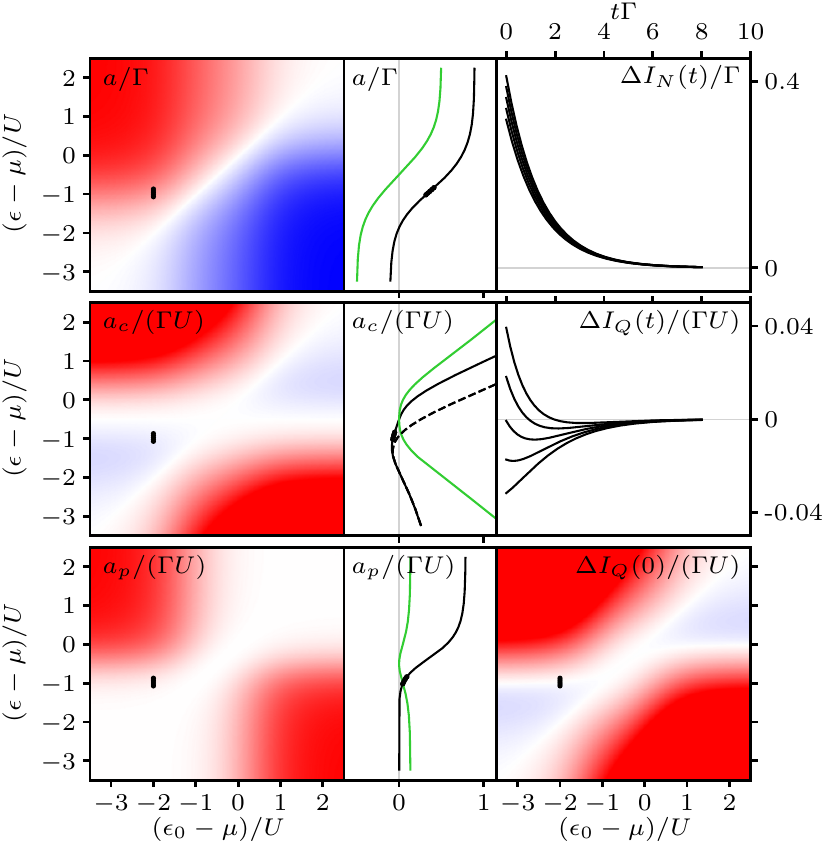}
	\caption{
		Same as Fig.~\ref{fig:amplitudes_no_pairing}
		but for intermediate temperature $T = U$.
		For higher temperatures $T \gg U$ the cooling effect stays (blue area in $\Delta I_Q(0)$)
		but all nonmonotonicity of the transients $I_Q(t)$ disappears (Sec.~\ref{sec:nonmonotonicity}).
	}
	\label{fig:nosuperconductor_reversal}.
\end{figure}

In Fig.~\ref{fig:nosuperconductor_reversal}
we illustrate how the no-superconductor results ($\alpha=0$) in Fig.~\ref{fig:no_interaction},
characteristic for $T \ll U$,
change when the thermal broadening becomes comparable to the interaction, $T \sim U$.
Note that although $U$ starts to be dominated by temperature here
it is crucial for the \emph{reversal effect} (transient heat current initially negative turns positive)
since for $U=0$ the transient heat current is single exponential and monotonically decaying.
Note however that for $U=0$ the transient heat current can still be negative \emph{all the time}
without reversing (not shown).
Interestingly, this is a cooling effect in the total heat current since the stationary heat current is zero in this single-terminal case.

%%%%%%%%%%%%%%%%%%%%%%%%%%%%%%%%%%%%%%%%%%%%%%%%%%%%%%%
%%%%%%%%%%%%%%%%%%%%%%%%%%%%%%%%%%%%%%%%%%%%%%%%%%%%%%%
%%%%%%%%%%%%%%%%%%%%%%%%%%%%%%%%%%%%%%%%%%%%%%%%%%%%%%%
%%%%%%%%%%%%%%%%%%%%%%%%%%%%%%%%%%%%%%%%%%%%%%%%%%%%%%%

% LEAVE FOLLOWING EMPTY LINE HERE FOR mksubmit script !
%apsrev4-2.bst 2019-01-14 (MD) hand-edited version of apsrev4-1.bst
%Control: key (0)
%Control: author (8) initials jnrlst
%Control: editor formatted (1) identically to author
%Control: production of article title (0) allowed
%Control: page (0) single
%Control: year (1) truncated
%Control: production of eprint (0) enabled
%
\end{document}